\documentclass[twocolumn]{aastex63}
\pdfoutput=1 
\usepackage[maxfloats=40]{morefloats}
\usepackage[T1]{fontenc}
\usepackage{color}
\usepackage{catoptions}
\usepackage{booktabs}
\usepackage{amsmath}
\newcommand{\Rom}[1]{\uppercase\expandafter{\romannumeral #1\relax}}

\begin{document}


\AuthorCallLimit=500
\title{Unveiling the importance of magnetic fields in the evolution of dense clumps formed at the waist of bipolar H\,{\sc{ii}} regions: 
a case study on Sh2-201 with JCMT SCUBA-2/POL-2}


\author[0000-0003-4761-6139]{Chakali Eswaraiah}
\affiliation{CAS Key Laboratory of FAST, National Astronomical Observatories, Chinese Academy of Sciences, Datun Road, Chaoyang District, Beijing 100101, People's Republic of China}
\affiliation{Institute of Astronomy and Department of Physics, National Tsing Hua University, Hsinchu 30013, Taiwan}
\correspondingauthor{Eswaraiah Chakali}
\email{eswaraiahc@nao.cas.cn,eswaraiahc@outlook.com}

\author[0000-0003-3010-7661]{Di Li}
\affiliation{CAS Key Laboratory of FAST, National Astronomical Observatories, Chinese Academy of Sciences, Datun Road, Chaoyang District, Beijing 100101, People's Republic of China}
\affiliation{University of Chinese Academy of Sciences (UCAS), Beijing 100049, People's Republic of China; dili@nao.cas.cn}
\email{dili@nao.cas.cn}

\author[0000-0002-9431-6297]{Manash R. Samal}
\affiliation{Physical Research Laboratory (PRL), Navrangpura, Ahmedabad 380 009, Gujarat, India; manash@prl.res.in}
\affiliation{Graduate Institute of Astronomy, National Central University 300, Jhongli City, Taoyuan County 32001, Taiwan}
\email{manash@prl.res.in}

\author[0000-0002-6668-974X]{Jia-Wei Wang}
\affiliation{Institute of Astronomy and Department of Physics, National Tsing Hua University, Hsinchu 30013, Taiwan}
\affiliation{Academia Sinica Institute of Astronomy \& Astrophysics, 11F of Astronomy-Mathematics Building, AS/NTU, No.1, Section 4, Roosevelt Road, Taipei 10617, Taiwan}

\author[0000-0002-8051-5228]{Yuehui Ma}
\affil{Purple Mountain Observatory and Key Laboratory of Radio Astronomy, Chinese Academy of Sciences, 10 Yuanhua Road, Nanjing 210033, People's Republic of China}
\affil{University of Chinese Academy of Sciences, 19A Yuquan Road, Shijingshan District, Beijing 100049, People's Republic of China.}

\author[0000-0001-5522-486X]{Shih-Ping Lai}
\affiliation{Institute of Astronomy and Department of Physics, National Tsing Hua University, Hsinchu 30013, Taiwan}

\author{Annie Zavagno}
\affiliation{Aix Marseille Univ, CNRS, CNES, LAM, Marseille, France}

\author[0000-0001-8516-2532]{Tao-Chung Ching}
\affiliation{CAS Key Laboratory of FAST, National Astronomical Observatories, Chinese Academy of Sciences, Datun Road, Chaoyang District, Beijing 100101, People's Republic of China}

\author[0000-0002-5286-2564]{Tie Liu}
\affiliation{Shanghai Astronomical Observatory, Chinese Academy of Sciences, 80 Nandan Road, Shanghai 200030, People's Republic of China}

\author[0000-0002-8557-3582]{Kate Pattle}
\affil{Institute of Astronomy, National Tsing Hua University (NTHU), 101, Section 2, Kuang-Fu Road, Hsinchu 30013, Taiwan, R.O.C}
\affil{Centre for Astronomy, School of Physics, National University of Ireland Galway, University Road, Galway, Ireland}

\author[0000-0003-1140-2761]{Derek Ward-Thompson}
\affil{Jeremiah Horrocks Institute, University of Central Lancashire, Preston PR1 2HE, UK}

\author{Anil K. Pandey}
\affiliation{Aryabhatta Research Institute of Observational Sciences (ARIES), Manora-peak, Nainital, Uttarakhand-state, 263002, India}

\author{Devendra K. Ojha}
\affiliation{Department of Astronomy and Astrophysics, Tata Institute of Fundamental Research, Homi Bhabha Road, Mumbai 400 005, India}

\received{receipt date}
\revised{revision date}
\accepted{acceptance date}

\shorttitle{Magnetic fields in Sh2-201}
\shortauthors{Eswaraiah et al.}

\begin{abstract}
    We present the properties of magnetic fields (B-fields) in two clumps (clump 1 and clump 2), located at the 
	waist of the bipolar H\,{\sc ii} region Sh2-201, based on JCMT SCUBA-2/POL-2 observations of 
	850 $\mu$m polarized dust emission. We find that B-fields in the direction of the clumps are 
	bent and compressed, showing bow-like morphologies, which we attribute to the feedback effect of the 
	H\,{\sc ii} region on the surface of the clumps. Using the modified Davis-Chandrasekhar-Fermi method 
	we estimate B-fields strengths of 266 $\mu$G and 65 $\mu$G for clump 1 and clump 2, respectively. 
	From virial analyses and critical mass ratio estimates, we argue that clump 1 is gravitationally bound 
	and could be undergoing collapse, whereas clump 2 is unbound and stable. We hypothesize that the 
	interplay between thermal pressure imparted by the H\,{\sc ii} region, B-field morphologies, and the various 
	internal pressures of the clumps (such as magnetic, turbulent, and gas thermal pressure), 
	has the following consequences: (a) formation of clumps at the waist of the H\,{\sc ii} region; 
	(b) progressive compression and enhancement of the B-fields in the clumps; (c) stronger B-fields will 
	shield the clumps from erosion by the H\,{\sc ii} region and cause pressure equilibrium between the 
	clumps and the H\,{\sc ii} region, thereby allowing expanding I-fronts to blow away from the 
	filament ridge, forming bipolar H\,{\sc ii} regions; and (d) stronger B-fields and turbulence 
	will be able to stabilize the clumps. A study of a larger sample of bipolar H\,{\sc ii} 
	regions would help to determine whether our hypotheses are widely applicable.
\end{abstract}


\section{Introduction}\label{sec:introd}

\subsection{H\,{\sc ii} region feedback and  magnetic fields}

Massive stars with mass $>$8 M$_\sun$ influence their surroundings 
via (a) energetic jets and outflows during their initial stages, 
(b) stellar winds, radiation pressure, and H\,{\sc{ii}}  regions  (which drive shocks and 
ionization fronts (I-fronts)) during their 
intermediate stages, and (c) supernova explosions at the end of their lives \citep[e.g.,][]{Tanetal2014,Motteetal2018}. 
These factors impact the second generation of stars through the resultant injection of 
energy and momentum into the ambient medium. 
Stellar feedback has two potential consequences~--~first by injecting the
turbulence into the cloud it stabilizes the cloud 
against its own gravity and maintains it in a state of quasi-static equilibrium \citep{Krumholzetal2005,KrumholzTan2007,Federrath2013}; and second, by triggering star formation it reduces 
the lifetime of a cloud to a few free-fall time scales \citep{Elmegreen2007,Dobbsetal2011,VazquezSemadenietal2009}.
These effects, which are known as negative and positive feedback, respectively, result in 
reduced or enhanced levels of star formation in a cloud. 
The key agents involved in the processes described above include magnetic fields (hereafter B-fields), turbulence, gravity, and 
H\,{\sc ii} region feedback. However the relative importance of B-fields in comparison to the other parameters, and the complex interplay between them is poorly understood. 

\citet{Deharvengetal2015} and \citet{Samaletal2018} identified several bipolar H\,{\sc ii} regions 
in our Galaxy using {\it Herschel} and {\it Spitzer} data analyses. 
They have suggested that such regions form due to the anisotropic expansion of H\,{\sc ii} region in flat- or 
sheet-like filamentary clouds, in accordance with recent 2D 
numerical simulations \citep{Wareingetal2017,Wareingetal2018}. 
In addition, \citet{Samaletal2018} found that the most massive and compact clumps are 
always formed at the waist the bipolar H\,{\sc ii} regions (see their 
Figure 3) showing the signatures of high-mass star formation. 
Since such clumps are possible sites of massive star and cluster formation, 
understanding the role of B-fields along with stellar feedback, turbulence, and gravity, as well as
the interplay among them holds a key to understand the star formation in such environments. 
While all other parameters can be relatively well constrained, B-fields are difficult to probe, 
quantify, and constrain. 

Dust grains are believed to be aligned with respect to the B-field orientation
via the ``radiative alignment torques (RAT)" mechanism \citep{Lazarian2007,Lazarianetal2015,Anderssonetal2015}.
The RAT model predicts that asymmetric, nonspherical dust grains rotate as a result of radiative torques imparted by their local radiation field, and so align themselves with their long axis perpendicular to the ambient
B-fields \citep{DolginovMitrofanov1976,DraineWeingartner1997,WeingartnerDraine2003,LazarianHoang2007a}.
The polarized thermal dust emission yields two quantities~--~the polarization fraction and the polarization position angle, which 
reveal polarized dust characteristics and the plane-of-sky 
component of the B-field morphology, respectively.

Several studies have attempted to probe B-fields in regions with stellar feedback using 
optical, near-infrared, and sub-millimeter polarization observations 
\citep[][]{PereyraMagalhaes2007,Wisniewskietal2007,Kusuneetal2015,ChenZhiweietal2017,Pattleetal2017b,LiuTieetal2018}.
These studies have demonstrated that initially weak B-fields become stronger as a
consequence of feedback-driven compression. 
These stronger B-fields play a crucial role in the formation and evolution of a variety of structures around H\,{\sc{ii}} regions.
\citet{Eswaraiahetal2017} have carried out NIR polarimetry towards RCW57A, a bipolar H\,{\sc ii} region hosting
a filament and dense clumps at the waist of H\,{\sc ii} region, and found that
B-fields are not only important to the formation and evolution of the filamentary cloud, 
but also strong enough to constrain the flows of expanding I-fronts to form the bipolar H\,{\sc ii} regions.
However, they could not probe the B-fields in the deeply embedded clumps under the influence of early stellar feedback, due to their heavy extinction. In this study, we probe B-fields in the dense clumps located at the waist of a 
geometrically simple bipolar H\,{\sc ii} region Sh2-201 (hereafter S201). 

\subsection{Description of Sh2-201}\label{sec:descri_sh201}

\begin{figure}
\resizebox{8.5cm}{8.3cm}{\includegraphics{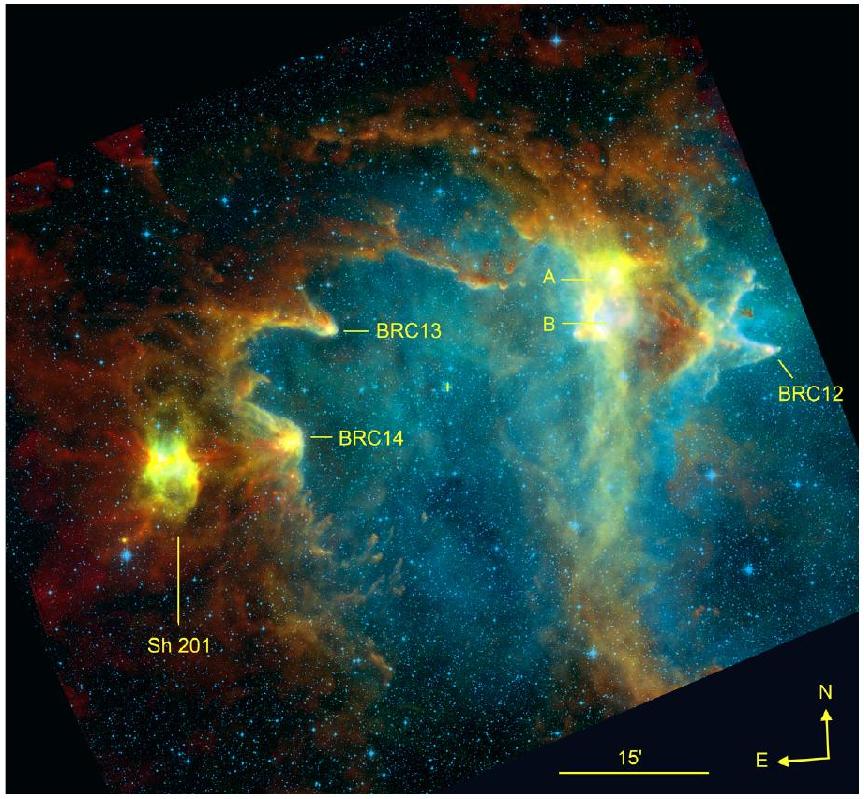}}
	\caption{The overall view of the W5-E complex and S201 region. Background image 
	is the color composite of {\it Herschel} SPIRE/250 $\mu$m (red; traces cold dust emission), 
	{\it Herschel} PACS/100 $\mu$m (green; traces warm dust, mainly from the 
	photodissociation regions (PDRs)), and DSS2-red survey (blue; traces H$\alpha$ emission) images. Various known 
	Bright Rimmed Clouds (BRCs) are shown. 
This figure is reproduced from Figure 2 of \citet{Deharvengetal2012} with permission.}
	\label{fig:fig2ofdeharvengetal2012}
\end{figure}

S201, with central coordinates of RA (J2000)$=$03$^{h}$03$^{m}$17$\fs$9, Dec (J2000)~$=$~$+$60$\degr$27$\arcmin$52$\arcsec$, 
is located to the east of the W5-E star-forming complex, as shown in Figure \ref{fig:fig2ofdeharvengetal2012}. 
This region is located at a distance of 2 kpc in the Perseus arm \citep{Megeathetal2008,Hachisukaetal2006}. 
It is a part of an elongated ($\sim$15$\arcmin$) filamentary cloud of mass 3.3$\times$10$^{4}$ M$_{\sun}$, as 
seen in $^{13}$CO \citep[see their Figure 1]{Niwaetal2009}. 
The local standard of rest velocities (V$_{\mathrm{LSR}}$) for the clumps of the W5-E region, as well as of S201, lie between 
$\sim-$38 km s$^{-1}$ and $\sim-$40 km s$^{-1}$ \citep{Niwaetal2009}. 
Similarly, the velocities of radio recombination lines (RRLs) of the ionized gas of S201 (V(RRLs)~$=$~$-$34.6 km s$^{-1}$, \citealt{Lockman1989}; 
V(H$\alpha$)~$=$~$-$35.5 km s$^{-1}$, \citealt{Fichetal1990}) are also in close agreement with those of the molecular gas of W5-E. 
Based on the distributions of (a) young stellar objects (Class 0, Class I, and Class II) from
{\it Spitzer} \citep{Koenigetal2008} and {\it Herschel} \citep{Deharvengetal2012} observations, (b) 
the physical conditions of the cold dust, (c) H\,{\sc{ii}} regions, and (d) exciting OB type stars, 
\citet{Deharvengetal2012} suggested that 
the entire W5-E complex and S201 are formed along a same parental dense, sheet-like, filamentary molecular cloud 
(see Figure \ref{fig:fig2ofdeharvengetal2012}). These results corroborate the hypothesis that S201 is a part of W5-E (see Figure \ref{fig:fig2ofdeharvengetal2012}).

\begin{figure*}
\centering
	\resizebox{16cm}{13cm}{\includegraphics{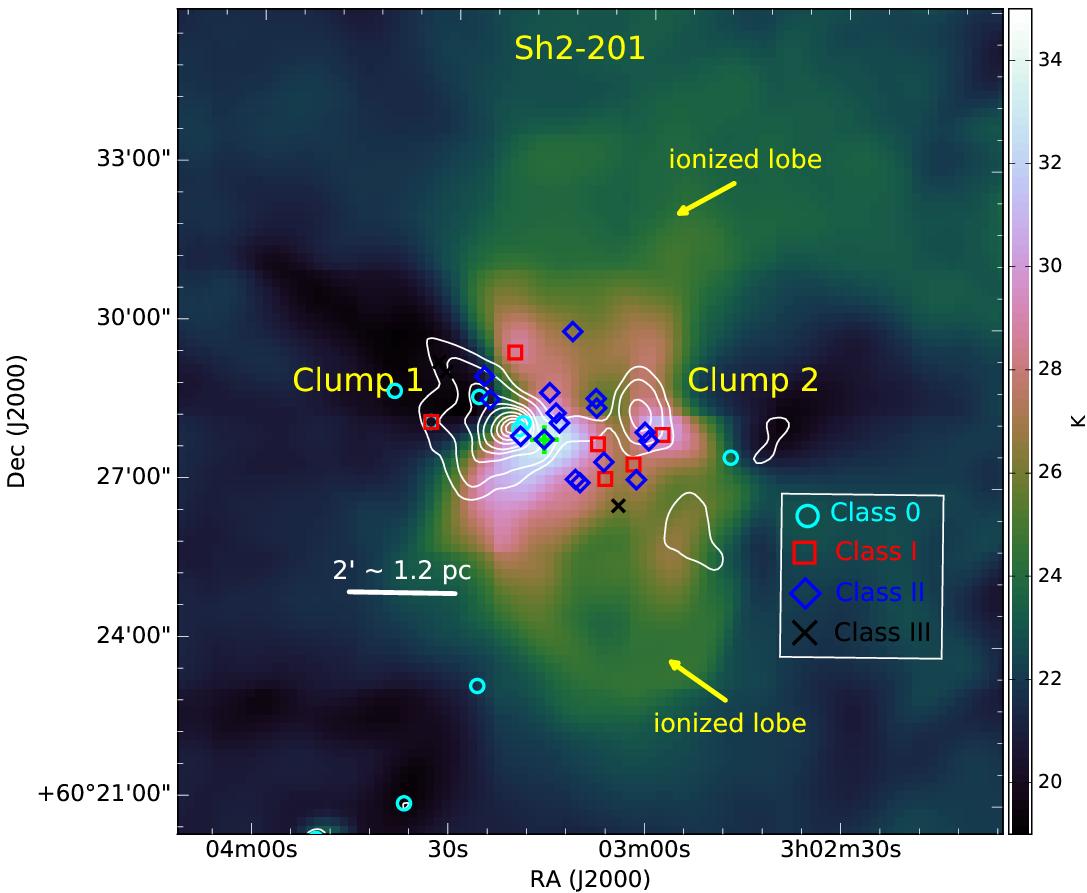}}
	\caption{The overall morphology and star formation activity in S201. Background image is the dust temperature map (in units of Kelvin). Column density map is overlaid using white contours at 
	levels of [4, 6, 12, 18, 24, 36, 48, 60, 72, 84]\% of the 
	peak column density of 6.94$\times$10$^{22}$ cm$^{-2}$. 
	The Class 0 sources, based on {\it Herschel} 100/160 $\mu$m data, are marked with 
	cyan circles \citep{Deharvengetal2012}. The Class I, II, and III sources, respectively, 
	are marked with square, diamond, and cross symbols \citep{Koenigetal2008}. 
	Positions of the two clumps, of swept-up matter around the two ionized lobes of the H{\sc ii} region, and of the ionizing source (green plus) are shown.
        Both dust temperature and column density maps are provided by \citet{Deharvengetal2012}. 
	A scale length of 2$\arcmin \sim$ 1.2\,pc is shown. \label{fig:tddustysos}}
\end{figure*}

Figure \ref{fig:tddustysos} shows a zoomed-in view of S201. 
NIR observations \citep{Ojhaetal2004} reveal that S201 hosts a compact embedded star cluster
containing more than one hundred stars and the most luminous member of the cluster 
is an O6--O8 zero age main sequence star (green plus symbol;  
Figures \ref{fig:tddustysos}, \ref{fig:polvecmaps}, and \ref{fig:diffpaBIG}(a)). 
As can be seen from Figure \ref{fig:tddustysos}, S201 is made up of two lobes extending from the 
center of the H\,{\sc{ii}} region and two dense clumps (namely, clump 1 and clump 2) at its waist. 
Radio, molecular hydrogen, and Br$\gamma$ images have revealed
arc-like or bow-like structured photo-dissociation regions at
the interface between the H\,{\sc{ii}} region and the clumps, 
highlighting the interaction between them \citep{Ojhaetal2004}. 
Several candidate Class 0 and Class I sources have been found within the vicinity of the clumps \citep{Koenigetal2008,Deharvengetal2012}. 
Since the life-time of Class 0/I YSOs is the order of 10$^{5}$ yr \citep[e.g.,][]{Evansetal2009},  
the clumps age is $\lesssim$10$^{5}$ yr. 
Therefore the clumps are likely to be in the early stages of their evolution, and so ideal candidates for 
investigating the interplay between B-fields, turbulence, gravity, and thermal pressure, and their 
implications for the formation and evolution of dense clumps and bipolar H\,{\sc ii} regions. 

This paper is organized as follows. Section \ref{sec:obs_red_anal} describes JCMT SCUBA-2/POL-2 observations, 
data reduction, and analyses. This section also presents molecular line ($^{13}$CO and C$^{18}$O) data from JCMT/HARP. 
Results based on the detailed analysis of B-field morphology and correlations between B-fields and 
intensity gradients (based on the VLA 21cm data) are 
presented in Section \ref{sec:results}. In this section, we also derive various parameters 
such as dust properties (gas column and number densities, and mass), gas kinematics (velocity dispersion and turbulent pressure), angular dispersion in B-fields (using structure function and auto-correlation function analyses), 
estimated B-field strength, and ionized gas properties (thermal and radiation pressure). 
Section \ref{sec:discuss} discusses the interplay between various parameters, stability analyses based on virial and critical mass estimates, and the consequences for the formation 
and evolution of clumps, and the formation of bipolar H\,{\sc ii} regions. 
Finally, the conclusions of our current study are 
summarized in Section \ref{sec:summary_conclusions}.

\section{Observations and data reduction}\label{sec:obs_red_anal}


\subsection{Dust continuum polarization observations using JCMT SCUBA-2/POL-2}\label{subsec:pol2obs}

Dust continuum polarization observations have been 
conducted using the POL-2 polarimeter installed on the SCUBA-2 camera (hereafter SCUBAPOL2) at the James Clerk Maxwell Telescope (JCMT; \citealt{Hollandetal2013}), a 15 m single dish submillimeter observatory located on the summit of Mauna Kea in Hawaii, USA. 
POL-2 observations of the S201 region (project code: M17BP041; PI: Eswaraiah Chakali) were 
carried out on 2017 November 18 using the POL-2 DAISY mapping mode \citep{Hollandetal2013,Fribergetal2016}. 
Three sets of observations were acquired under JCMT Band 1 
weather conditions, during which the atmospheric optical depth at 225 GHz, $\tau_{225}$, was 0.03. 
Each set was observed for 30 minutes, resulting in a total integration time of $\sim$1.5 hr. 

The POL-2 DAISY scanning mode observes a fully sampled 
circular region of 15 arcmin diameter. The rms noise 
is uniform within the central 3$\arcmin$-diameter region of the DAISY map, increasing 
towards the outer parts of the map. POL-2 data were taken at 450 and 850 $\mu$m simultaneously, 
with a resolution of 9$\farcs$6 
arcsec and 14$\farcs$1, respectively. Here we present the results of 850 $\mu$m data only, 
due to the low sensitivity of the 450 $\mu$m data. 
A flux calibration factor (FCF) of 725 Jy pW$^{-1}$ beam$^{-1}$ was applied 
to the 850 $\mu$m Stokes I, Q, and U parameters.   
This FCF value was derived by multiplying the typical SCUBA-2 FCF of
537 Jy pW$^{-1}$ beam$^{-1}$ \citep{Dempseyetal2013} by a transmission
correction factor of 1.35 measured in the laboratory and
confirmed empirically by the POL-2 commissioning team using 
observations of Uranus \citep{Fribergetal2016}.

The POL-2 data were reduced using the 
\textsc{Starlink} \emph{pol2map}\footnote{\url{http://starlink.eao.hawaii.edu/docs/sc22.pdf}} 
routine (adapted from the SCUBA-2 data reduction procedure \emph{makemap}), 
which was recently added to the \textsc{Starlink Smurf} mapmaking software suite 
\citep{Berryetal2005,Chapinetal2013,Currieetal2014}. 
To correct for the instrumental polarization at JCMT/850 $\mu$m, we employed the 
2018 January IP model during the data reduction, which was extensively tested by the POL-2
commissioning team \citep{Fribergetal2016,Fribergetal2018}. 
Instrumental polarization is typically $\sim$1.5\% of the measured total 
intensity \citep{Fribergetal2018}. 
More details on the equations and procedures used to derive the polarization
measurements: the debiased degree of polarization [$P$ (\%)], polarization angles [$\theta$ ($\degr$)],
Stokes parameters [$Q$ (\%) and $U$ (\%)], intensity [$I$ (mJy/beam)], and polarized intensity [$PI$ (mJy/beam)] along with their 
uncertainties, can be found in recently-published work \citep[and references therein]{WangJWetal2019,Coudeetal2019,LiuJunhaoetal2019}. 

The final Stokes I, Q, and U maps have a pixel size of 4$\arcsec$; however, 
the polarization vector catalog was created by setting the {\it bin size} 
parameter in the third step of {\it pol2map} 
to 12$\arcsec$, in order to achieve better sensitivity. 
The mean rms noise in the Stokes I measurements, $\sigma_{I}$, is $\sim$5 mJy/beam (note that the mean rms noise 
in the $4\arcsec$ pixel-size Stokes I map is $\sim$14 mJy/beam). 
In order to infer the B-field orientation in the clumps, we have excluded the data corresponding to fainter
regions whose polarization measurements are generally noise-dominated.
Therefore, we adopted the following data selection criteria:
ratio of intensity to its uncertainty, $I/\sigma_{I}$,~$>$~10; and 
ratio of polarization fraction to its uncertainty, $P/\sigma_{P}$,~$>$~2, yielding a total of 62 polarization measurements, 
which are listed, along with their coordinates, in Table \ref{tab:polmeasurements}. 
We also listed $I$ and $PI$ along with their uncertainties.
It should be noted that the $\theta$ values listed have a correction of 90$\degr$,
and hence infer the B-field orientation\footnote{0$\degr$ corresponds to equatorial North and increases towards the 
East as per the IAU convention.} projected on the plane of sky. 

In this work, we investigate the morphology and strength of B-fields. 
Results based on the polarization characteristics and alignment efficiency of the dust grains
will be published elsewhere (Eswaraiah et al. in prep), 
and will consist of various analyses of the 
relationship between $P$ and $I$ using the POL2 data of S201.

\subsection{Molecular lines data from JCMT HARP}\label{subsec:molecularjcmtharp}

The JCMT is also equipped with the Heterodyne Array
Receiver Program (HARP)/Auto-Correlation Spectral Imaging
System (ACSIS) high-resolution heterodyne spectrometer, 
capable of observing molecular lines between 325 and 375 GHz (or 0.922 mm and 0.799 mm). 
HARP is a 4 $\times$ 4 detector array that can be used in combination with ACSIS
to rapidly produce large-scale velocity maps of astronomical
sources \citep{Buckleetal2009}. In this paper, we use archival 
$^{13}$CO (3--2) and C$^{18}$O (3--2) molecular line data ($\sim$14$\arcsec$ resolution, 
project ID: M09BU04, PI: Mark Thompson, observed on 2009-08-25)
to examine the distribution of gas and to extract the gas velocity dispersion values in clumps 1 and 2.

\section{Analyses and Results}\label{sec:results}

\subsection{B-field morphology}

\begin{figure*}
\centering
\resizebox{18cm}{9cm}{\includegraphics{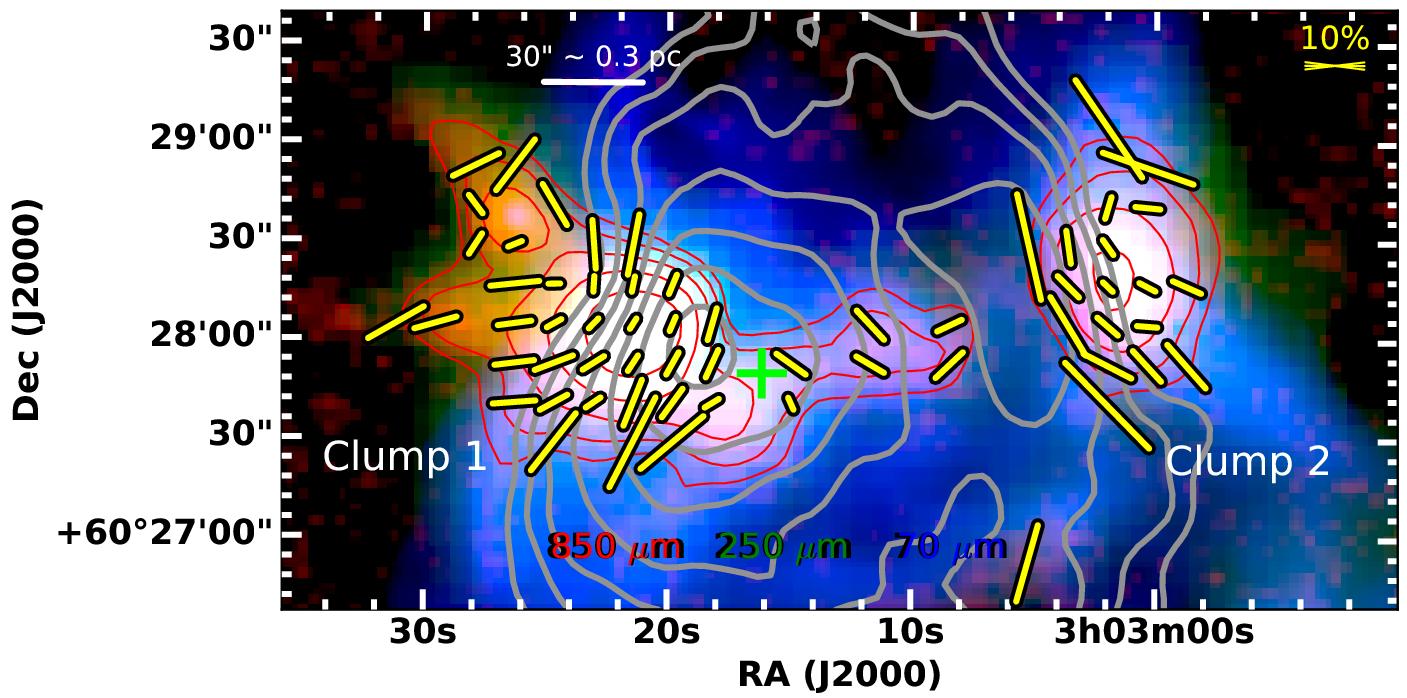}}
\resizebox{18cm}{9cm}{\includegraphics{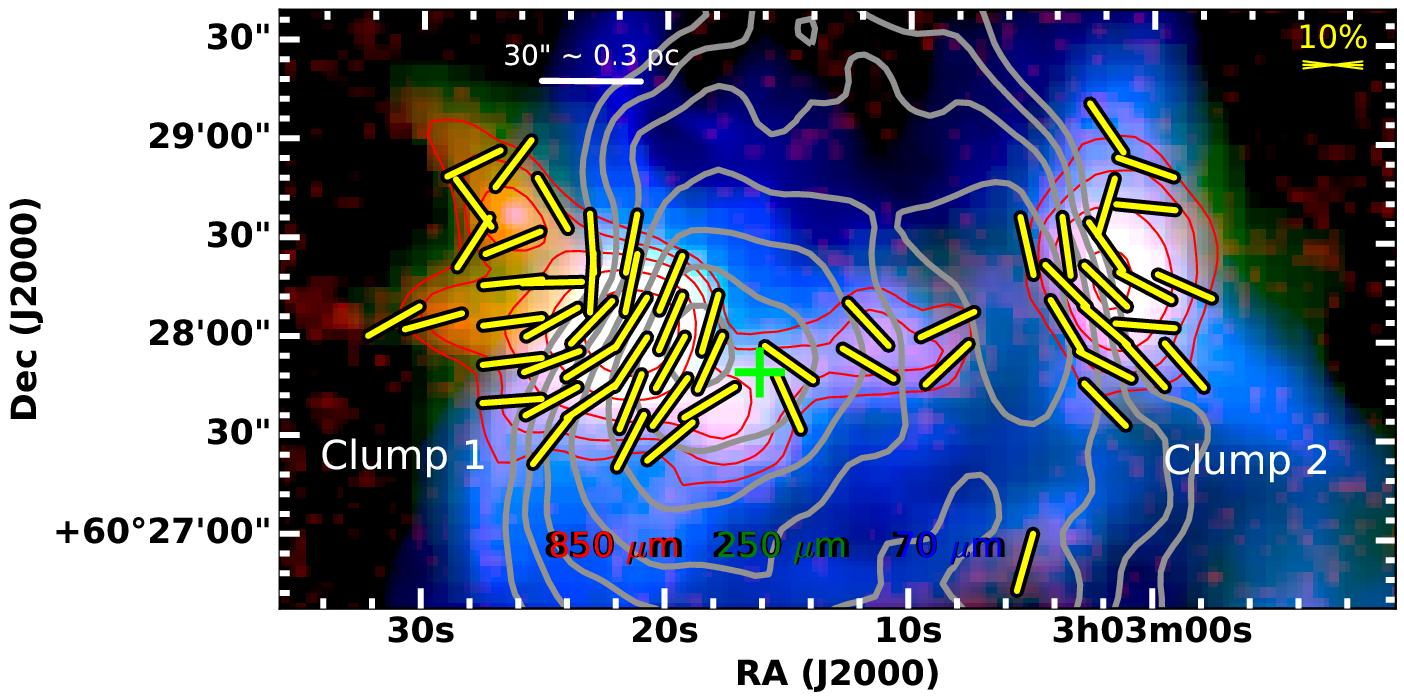}}
	\caption{Vector maps showing B-field orientations, 
	lengths proportional to polarization fraction (top panel) and with fixed lengths (bottom panel). 
	B-field vector maps are overlaid on the color composite of JCMT/SCUBAPOL2 850 $\mu$m Stokes I (red), 
	{\it Herschel} SPIRE/250 $\mu$m (green), and {\it Herschel} PACS/70 $\mu$m (blue) images. 
	Red contours correspond to JCMT/SCUBAPOL2 850 $\mu$m Stokes I map and are drawn at 
	$[$3, 6, 12, 24, 48, 96, 192$]$ $\times$ the rms noise of 14 mJy/beam. 
	Gray contours, corresponding to the VLA/21 cm continuum emission representing the distribution of 
	the ionized medium of H\,{\sc{ii}} region, are drawn at $[$1, 3, 6, 12, 24, 48, 96, 206$]$ $\times$ the rms noise of 2.3$\times$10$^{-4}$ mJy/beam (where beam size $\sim17\arcsec\times13\arcsec$). In both panels, reference vectors with a B-field orientation of 90$\degr$, along with mean uncertainty of 7$\degr$ are shown.}\label{fig:polvecmaps}
\end{figure*}

The measured $P$ values range from $\sim$2\% to $\sim$25\% with a
mean and standard deviation of $\sim$7$\pm$5\%, while 
the B-field orientations ($\theta$) range from $\sim$4$\degr$ to $\sim$177$\degr$ with a mean and 
standard deviation of $\sim$99$\pm$50$\degr$; a higher standard deviation implies a widely distributed B-field morphology with 
multiple components. The mean measured uncertainties in $P$ and $\theta$ are
$\sim$2$\pm$1\% and $\sim$7$\pm$4$\degr$, respectively. 
The mean uncertainties in Stokes parameters, $\sigma_{I}$, $\sigma_{Q}$, and $\sigma_{U}$, are found 
to be $\sim$5, $\sim$2, and $\sim$2 mJy beam$^{-1}$, respectively. 
Similarly, the mean uncertainties in polarization ($\sigma_{P}$) and B-field orientation ($\sigma_{\theta}$) 
are found to be $\sim$2\% and $\sim$7$\degr$, respectively.

Our aim is to derive various parameters for the two clumps at the waist of Sh201. We thus 
separate the polarization data according to the areas covered by individual clumps, and by the ionized medium. 
Of the 62 total measurements, we find that 36 and 18 are in the direction of clumps 1 and 2, respectively, while the remaining 8 measurements are located between
or away from the two clumps, and are excluded assuming that they are not representatives of either clump.

The B-field geometry, based on our 62 measurements, is superimposed on the color composite of POL-2 Stokes I, 
{\it Herschel}/SPIRE 250 $\mu$m, and {\it Herschel}/PACS 70 $\mu$m images shown in Figure \ref{fig:polvecmaps}. 
Red and gray contours correspond to the distributions of 
dust emission (based on POL-2 850 $\mu$m I map) and of the H\,{\sc{ii}} region (based on 
VLA 1.45 GHz/21~cm continuum\footnote{VLA 21~cm image is downloaded from 
\url{https://archive.nrao.edu/archive/advquery.jsp}}), respectively. 
Evidently, B-fields in clump 1 follow a bow-like morphology and are conspicuously compressed 
at the interface region between the dust emission and the ionized medium. This interaction can be 
witnessed from the closely spaced 21-cm contours. 
B-fields also seem to be compressed in clump 2 but with a lower degree of curvature. 

\begin{figure}
\resizebox{8.75cm}{6.75cm}{\includegraphics{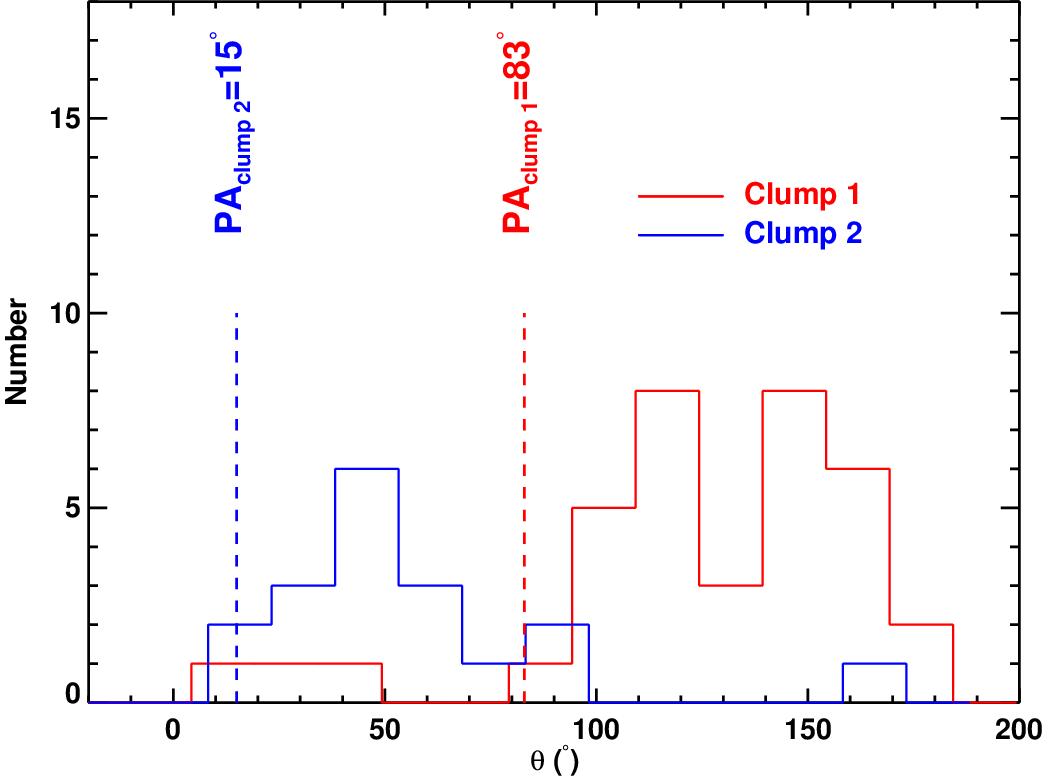}}
	\caption{Distributions of B-field orientation of clumps 1 and 2, 
	represented by red and blue histograms, 
	respectively. Position angles of the major axes of clumps 1 and 2 are respectively shown 
	with dashed red and blue lines at 91$\degr$ and 8$\degr$.}\label{fig:histpa}
\end{figure}

Histograms of the B-field orientation, shown in Figure \ref{fig:histpa}, reveal
 the existence of two major components in clump 1. 
One component, located near the interaction region of the H\,{\sc{ii}} region and the dust emission (POL2 Stokes I), 
peaks at $\sim$150$\degr$ and is oriented northwest-southeast. The other component, located on the eastern side of clump 1, is oriented at $\sim$115$\degr$, approximately along the east-west direction, nearly parallel to the major axis (position angle $\sim$83$\degr$) of clump 1. Conversely, the B-fields in clump 2 exhibit a single component with a prominent peak at $\sim$50$\degr$ and is oriented northeast-southwest. This B-field component is neither parallel nor perpendicular to 
the major axis (position angle of $\sim$15$\degr$) of clump 2. Figures \ref{fig:polvecmaps} and \ref{fig:histpa} imply the presence of multiple B-field components in clumps 1 and 2. 

\subsection{Intensity (ionized gas) gradients versus B-fields}\label{subsec:bintgrad}

\begin{figure*}
\centering
\resizebox{18cm}{11cm}{\includegraphics{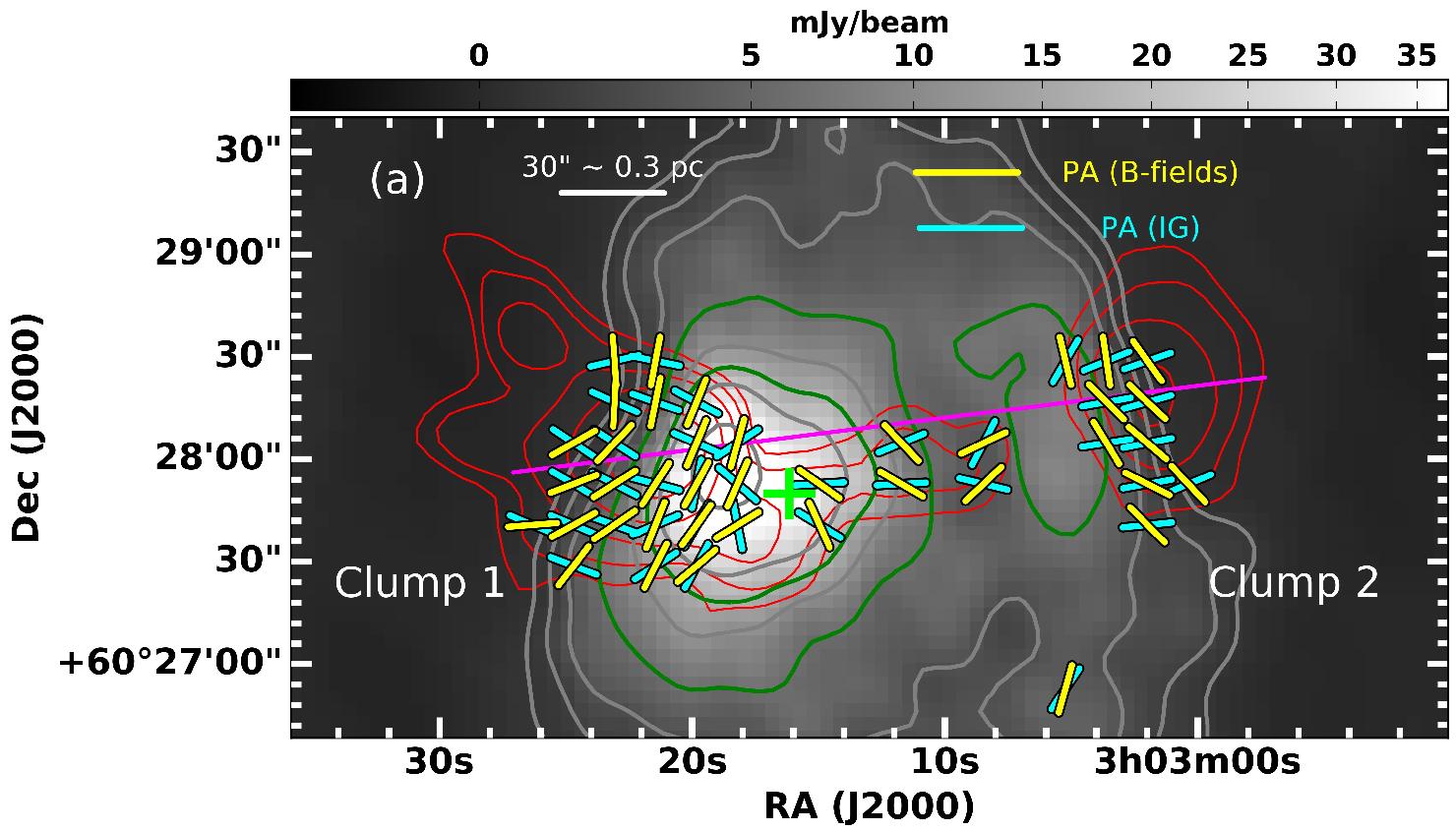}}
	\resizebox{8.5cm}{7.4cm}{\includegraphics{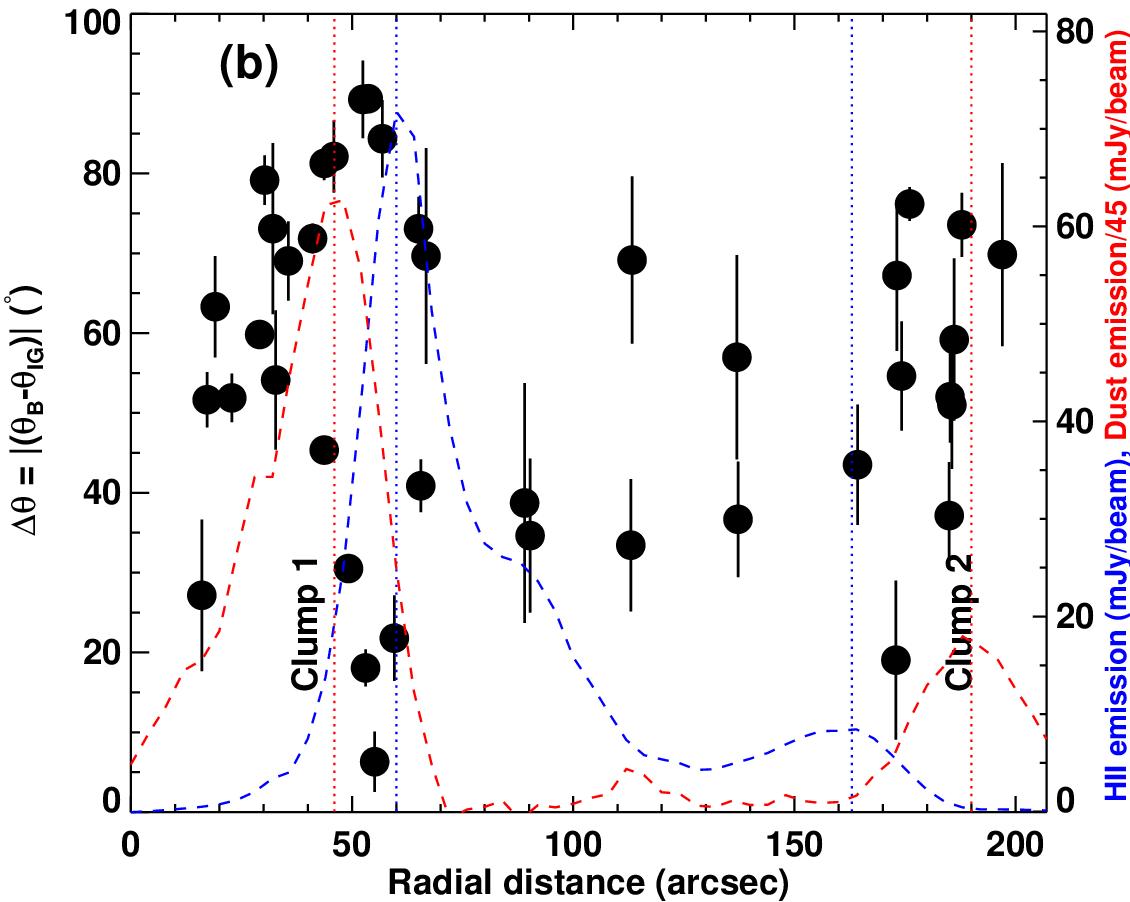}}
	\resizebox{8.5cm}{7.4cm}{\includegraphics{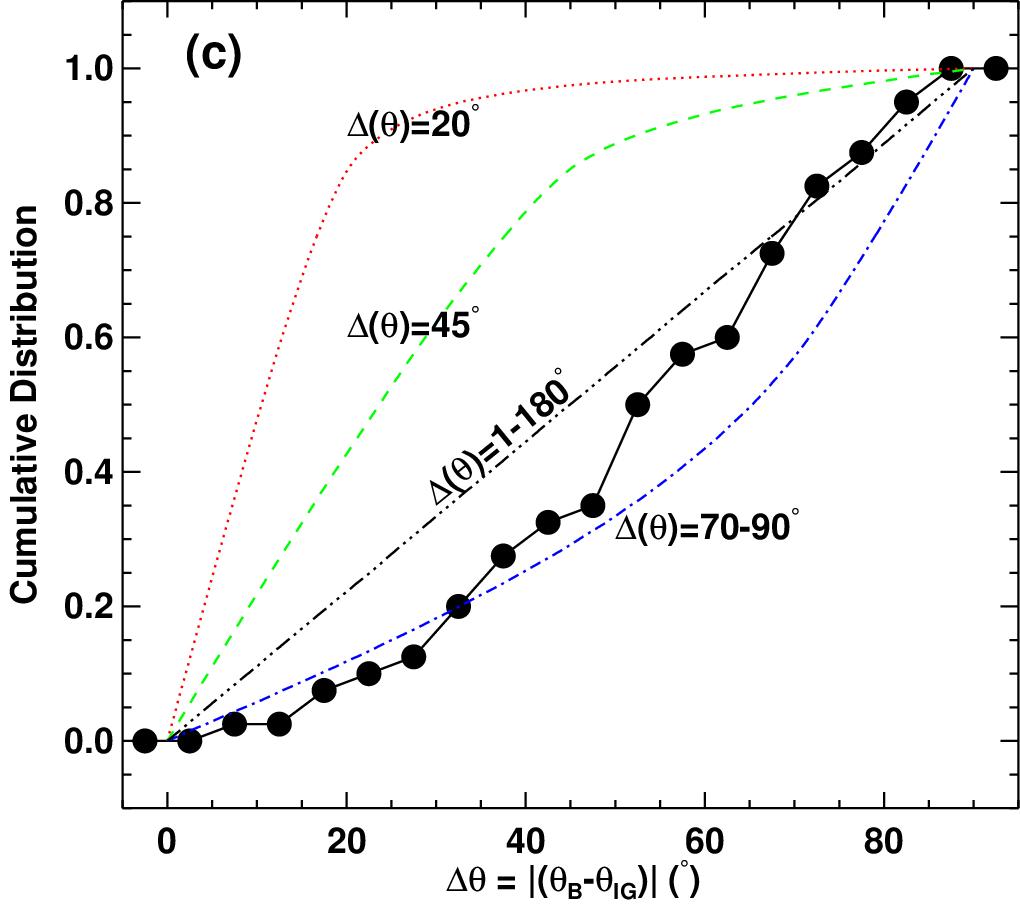}}
\caption{(a) The orientations of B-fields ($\theta_{B}$; yellow vectors) and Intensity Gradients ($\theta_{IG}$; cyan vectors) are overlaid on the VLA 1.45 GHz continuum map. 
Red and gray contours are the same as those shown in Figure \ref{fig:polvecmaps}. 
Integrated fluxes within the green contours correspond to 26$\sigma$ and 70$\sigma$ 
flux levels of 0.006 Jy/beam and 0.016 Jy/beam around clumps 1 and 2, respectively, 
are used for estimating the thermal pressures exerted on the respective 
clump surfaces (c.f. Section \ref{subsec:radio_ptherm} for more details). 
Locations of the clumps and reference scale length are also shown. 
Radial profiles of dust and H\,{\sc ii} emission are extracted 
along the magenta line and are drawn in panel (b). 
(b) The offset between the position angles of B-fields and intensity gradients, i.e.,
        $\Delta\theta$~$=$~$\vert(\theta_{B}-\theta_{IG})\vert$ as a function of radial distance (filled circles). 
        The zero radial distance points to the left edge of the magenta line close to the clump 1. 
        Also, in the right-hand of y-axis, we plotted the radial variation of radio emission 
	(VLA 21\,cm; blue dashed lines) as well as of the dust emission 
	(SCUBAPOL2 Stokes I; red dashed line) along the same magenta line shown in panel (a).
        (c) Cumulative distribution of $\Delta\theta$~$=$~$\vert(\theta_{B}-\theta_{IG})\vert$ is shown with filled circles. Model cumulative distributions (lines with different colors) for $\Delta\theta$ values 
	of 20$\degr$, 45$\degr$, 70-90$\degr$, and random angles, 
	adopted from Monte Carlo simulations \citep{Hulletal2014}, are also shown.}
        \label{fig:diffpaBIG}
\end{figure*}

In order to examine whether the multiple B-field components of S201 are shaped by the expanding I-front, 
we construct intensity gradients using the 
VLA 21\,cm continuum intensity map. 
More details on making the intensity gradient map 
are given in Appendix \ref{sec:IngGrads_appendix}. 
To compare the orientation of the intensity gradients ($\theta_{\mathrm{IG}}$) 
with those of the B-fields ($\theta_{\mathrm{B}}$), 
we estimate mean $\theta_{\mathrm {IG}}$ over a $\sim$14$\arcsec$ diameter region 
(corresponding to the beam size of JCMT) 
around each $\theta_{\mathrm{B}}$ vector. Figure~\ref{fig:diffpaBIG}(a) 
shows the pairs of $\theta_{\mathrm{IG}}$ (cyan vectors) 
and $\theta_{\mathrm{B}}$ (yellow vectors) overlaid on the VLA 21\,cm continuum intensity map. 

Figure \ref{fig:diffpaBIG}(b) shows the offset angle ($\Delta\theta$) 
between $\theta_{\mathrm{B}}$ and  $\theta_{\mathrm{IG}}$ 
as a function of radial distance (from clump 1 to clump 2) 
along the magenta line shown in Figure \ref{fig:diffpaBIG}(a). 
The radial profiles of dust and H\,{\sc ii} emission, extracted along the magenta line, are 
also shown to examine their correlation with the $\Delta\theta$ values. 
At radial distances $<$ 70$\arcsec$ and $>$ 160$\arcsec$, 
where the dust and H\,{\sc ii} emission peaks of clump 1 and 2, respectively, are found, 
the majority of the $\Delta\theta$'s lie between $\sim$50$\degr$ and $\sim$90$\degr$. 
Notably, perpendicular components prevail 
around clump 1, and also at clump 2, albeit with less prominence due to the lack of $\Delta\theta$ 
values in the range 70$\degr$ to 90$\degr$. 
Here, $\Delta\theta$~$=$~90$\degr$ refers to perpendicular alignment of B-fields with 
intensity gradients or parallel alignment of B-fields with intensity contours.
Between clumps 1 and 2, i.e., from $\sim80\arcsec$ to $\sim150\arcsec$, the $\Delta\theta$ values 
are neither parallel nor perpendicular, lying between 
$\sim30\degr$ and $\sim40\degr$. These values contribute towards the 
random component as evident from Figures \ref{fig:diffpaBIG}(b) and (c). 
In addition, there also exist a few parallel components ($\sim0\degr$~--~$\sim30\degr$) near clump 1 and clump 2. 

Therefore, at clump 1, because of the prominent interaction between the dust and H\,{\sc{ii}} emissions, 
B-fields tend to be perpendicular to the intensity gradients ($\theta_{\mathrm{IG}}$) or parallel to the 
intensity contours. Whereas at clump 2, as the interaction between the dust and the H\,{\sc ii} region is less prominent 
(based on their peak brightness in 
comparison to those of clump 1), a lower degree of alignment between B-fields and 
intensity contours is evident. 
The cumulative distribution of $\Delta{\theta}$ values also confirms 
the possibility of both perpendicular and random components, as shown in 
Figure \ref{fig:diffpaBIG}(c). In order to study the projection effect from 
alignments between B-fields and intensity gradients in three-dimensional space to 
those in the plane of the sky, we also show cumulative distributions based on 
Monte Carlo simulations \citep{Hulletal2014}. These simulations randomly select pairs of orientations in three 
dimensions that have alignments in the ranges 0$\degr$~--~20$\degr$, 0$\degr$~--~45$\degr$, 70$\degr$~--~90$\degr$, or random alignment; then their $\Delta{\theta}$ are measured on the plane of the sky. The resulting cumulative distribution functions of the simulations are shown in Figure \ref{fig:diffpaBIG}(c).
Kolmogorov-Smirnov statistics suggest that, with a probability of 82.5\%, 
our data is consistent with the model distribution corresponding to a $\Delta{\theta}$ of 70~--~90$\degr$, 
while there is a probability of only 26.5\% that our data have a random component. 
Therefore, we conclude that the B-fields in the clumps are shaped by the H\,{\sc ii} region.

\subsection{Dust properties of the clumps: column density, number density, and mass}\label{subsec:cold_numb_mass} 

For an idealized cloud, the dust emission at frequencies where optical depth is small can be described by 
(\citealt{Hildebrand1983}; see also \citealt{LiDietal1999}) 
\begin{equation}
S(\nu) = N(\sigma/D^{2})Q(\nu)B(\nu,T_{d}),
\end{equation}
where $S(\nu)$ is the flux density (erg s$^{-1}$ cm$^{-2}$ Hz$^{-1}$ or Jy) from a cloud at distance $D$, $N$ is the number of spherical 
grains included in the cloud volume subtended by the beam, $Q(\nu)$ is the dimensionless absorption coefficient, $\sigma$ is the geometric 
cross section of a dust grain, and $B(\nu,T_{d})$ is the Planck function for a blackbody at temperature $T_{d}$.

The above equation can be written as follows to construct 
the column density map from the POL2 850 $\mu$m Stokes I map \citep{Kauffmann2007}, 
	\begin{eqnarray*}\label{eq:coldesti}
	N_{\mathrm{H_{2}}} = 2.02 \times 10^{20} \mathrm{cm^{-2}} \left(e^{1.439{\left(\frac{\lambda}{\mathrm {mm}}\right)}^{-1} 
	{\left(\frac{\mathrm{T_{d}}}{\mathrm{10 K}}\right)}^{-1}} - 1\right) \\
	{\left(\frac{\kappa_{\nu}}{0.01\,{\mathrm {cm^{2}\,g^{-1}}}}\right)}^{-1}  \left(\frac{{S_{\nu}}^{\mathrm {beam}}}{\mathrm {mJy\,beam^{-1}}}\right) {\left(\frac{\theta_{\mathrm{HPBW}}}{\mathrm{10\,arcsec}}\right)}^{-2} {\left(\frac{\lambda}{\mathrm {mm}}\right)}^{3}, \\
\end{eqnarray*}
where $T_{d}$ is the mean dust temperature within the two clumps, $\lambda$~$=$~0.85~mm, 
and $\theta_{\mathrm{HPBW}}$~$=$~beam size (14$\arcsec$), $\kappa_{\nu} = 0.1 (\nu / 1 \mathrm{THz})^{\beta} = 0.0182$ is the dust opacity 
in cm$^2$ g$^{-1}$, and $\beta$ is the dust opacity exponent of 2 (e.g., \citealt{Arzoumanianetal2011}). 

We have performed CASA 2D Gaussian fits on the POL2 850 $\mu$m Stokes I map, 
specifically on the pixels around each clump 
having I $>$ 140 mJy/beam (i.e., $>$10$\sigma$, where $\sigma$ = 14 mJy beam$^{-1}$ is the rms noise 
in the I map with pixel size~$=$~4$\arcsec$), 
and extracted the spatial extents of the clumps. The resultant dimensions of the clumps, 
along with their central coordinates, are given in Table \ref{tab:paramscl12}. 
The effective radius of each clump is $R_{\mathrm{eff}}$~$=$~$\sqrt{\sigma_{a} \times \sigma_{b}}$; 
where $\sigma_{a}$ and $\sigma_{b}$ are the extents of the semimajor and semiminor axes, respectively, 
and are found to be 13.3$\pm$0.3$\arcsec$
(or 0.13$\pm$0.01~pc) for clump 1 and 15.2$\pm$0.4$\arcsec$ (or 0.15$\pm$0.01~pc) for clump 2.
The mean dust temperatures ($T_{d}$), within the dimensions of clump 1 and clump 2, are 
estimated to be 27 K and 29 K \citep{Deharvengetal2012}, and 
are used in the above Equation to estimate the respective column density maps. 

The total column densities ($\sum N_{H_{2}}$) are estimated within the clump dimensions, and 
are found to be (9.2$\pm$1.6)$\times$10$^{23}$ cm$^{-2}$ for 
clump 1 and (3.5$\pm$0.5)$\times$10$^{23}$ cm$^{-2}$ for clump 2.
The number density ($n(H_{2})$) is estimated using the relation 
\begin{equation}
n(H_{2}) = \frac{\sum\,N_{H_{2}}\,\times\,A_{\mathrm{pixel}}}{{\frac{4}{3}}{{\pi\,R_{\mathrm{eff}}}^{3}}}, 
\end{equation}
where $A_{pixel}$ is the area of a pixel (4$\arcsec$) in cm$^{2}$. 
$R_{eff}$ is the effective radius (estimated above). 
The derived number densities for clumps 1 and 2, respectively, are 
(5.1$\pm$0.9)$\times$10$^{4}$~cm$^{-3}$ and (1.3$\pm$0.2)$\times$10$^{4}$~cm$^{-3}$. The mean 
column density ($N_{H_{2}}$) for clump 1 is (27$\pm$6)$\times$10$^{21}$ cm$^{-2}$ and for clump 2 is 
(8$\pm$1)$\times$10$^{21}$ cm$^{-2}$.

We have estimated clump mass using the relation 
\begin{equation}
M~=~A_{pixel}\,\mu_{H_{2}}\,m_{H}\,\sum N_{H_{2}}.
\end{equation}
Here we used the integrated total column densities, $\sum N_{H_{2}}$, within the contours corresponding 
to 10$\sigma$ Stokes I. 
The resulting masses are found to be 72$\pm$5 M$_{\sun}$ 
and 22$\pm$2 M$_{\sun}$ for clumps 1 and 2, respectively.

We note that the masses determined above are likely to be lower limits on
 the true masses because they have been estimated using
the average dust temperature over the clump areas. 
We thus measured masses within the areas corresponding to 10$\sigma$ Stokes I but from the column density map 
constructed from the {\it Herschel} temperature map
(for details see \citealt{Deharvengetal2012}). The resulting masses are found to 
be 191$\pm$13 M$_{\sun}$ and 30$\pm$3 M$_{\sun}$ for clump 1 and clump 2, respectively. 
Although these values agree within a factor of two with 
the masses derived from 850 $\mu$m Stokes I map, they are likely to better represent
 the true masses. We thus used these values for further analyses. The number 
densities and masses estimated above are listed in Table \ref{tab:paramscl12}. 

\subsection{Gas properties: velocity dispersion}\label{subsec:gasveldisp}

\begin{figure}
\resizebox{7.75cm}{9.75cm}{\includegraphics{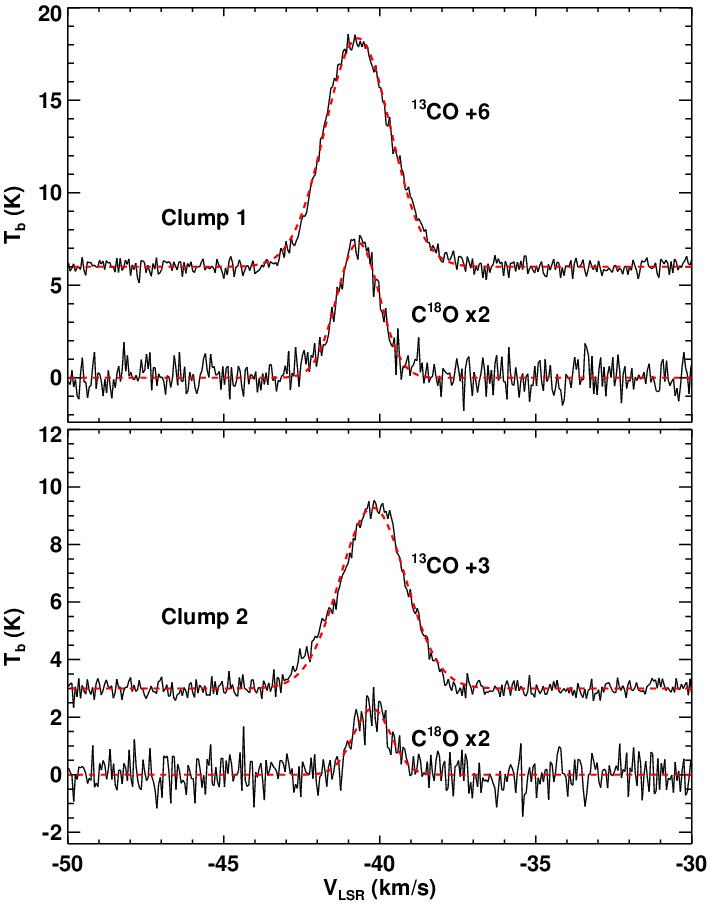}}
	\caption{$^{13}$CO (3--2) and C$^{18}$O (3--2) brightness temperature versus $V_{\mathrm{LSR}}$ spectra, 
	averaged over the extents of clumps 1 (top) and 2 (bottom). To view clearly, the 
	spectra of $^{13}$CO and C$^{18}$O are shifted by adding and multiplying by arbitrary numbers as shown in 
	the figure. Best-fit Gaussians are shown with red dashed lines.}\label{fig:13COandC18Ospectra}
\end{figure}

Figure \ref{fig:13COandC18Ospectra} shows the 
JCMT HARP/ACSIS $^{13}$CO(3--2) and C$^{18}$O(3--2) spectra of the two 
clumps, averaged over the clump dimensions, in units of brightness temperature ($T_{b}$, K) 
versus velocity (V$_{\mathrm{LSR}}$, km s$^{-1}$).
We performed Gaussian fitting on each spectrum,
resultant peak brightness temperature ($T_{b,p}$), central V$_{\mathrm{LSR}}$, and velocity 
dispersion ($\sigma_{V_{\mathrm{LSR}}}$) values are given in Table \ref{tab:gaussfitresults}.
The $\sigma_{V_{\mathrm{LSR}}}$ values for clumps 1 and 2 are 1.05$\pm$0.01 km s$^{-1}$ 
and 1.06$\pm$0.01 km s$^{-1}$ in $^{13}$CO, and 0.69$\pm$0.02 km s$^{-1}$ and 0.60$\pm$0.04 km s$^{-1}$ in C$^{18}$O.
For clump 1, using C$^{18}$O (J=1--0) measurements, \citet[][clump 9 in their work]{Niwaetal2009} 
derived the mean velocity dispersion over an area somewhat larger than that we consider 
as 0.71 km s$^{-1}$, which is in close agreement with our estimate derived from C$^{18}$O(3--2). 

\begin{figure*}
\centering
\resizebox{8.75cm}{7.8cm}{\includegraphics{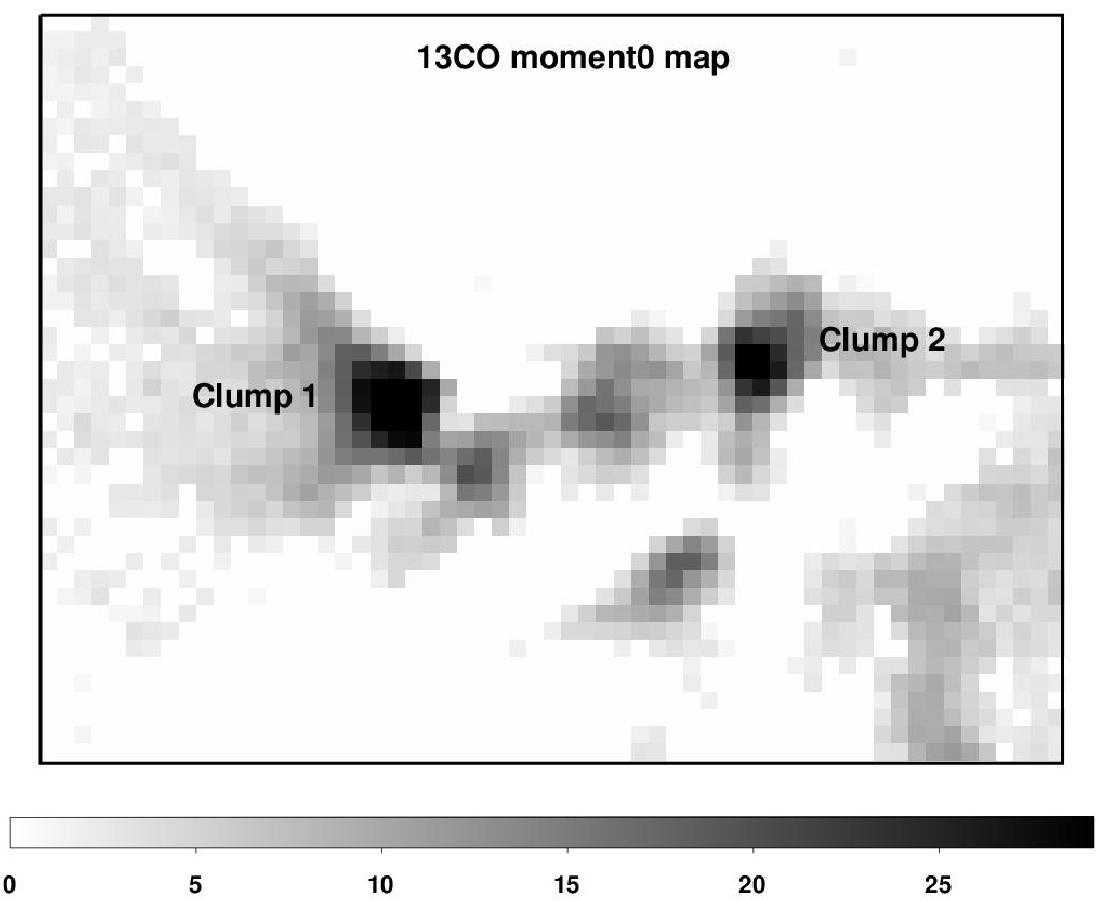}}
\resizebox{8.75cm}{7.8cm}{\includegraphics{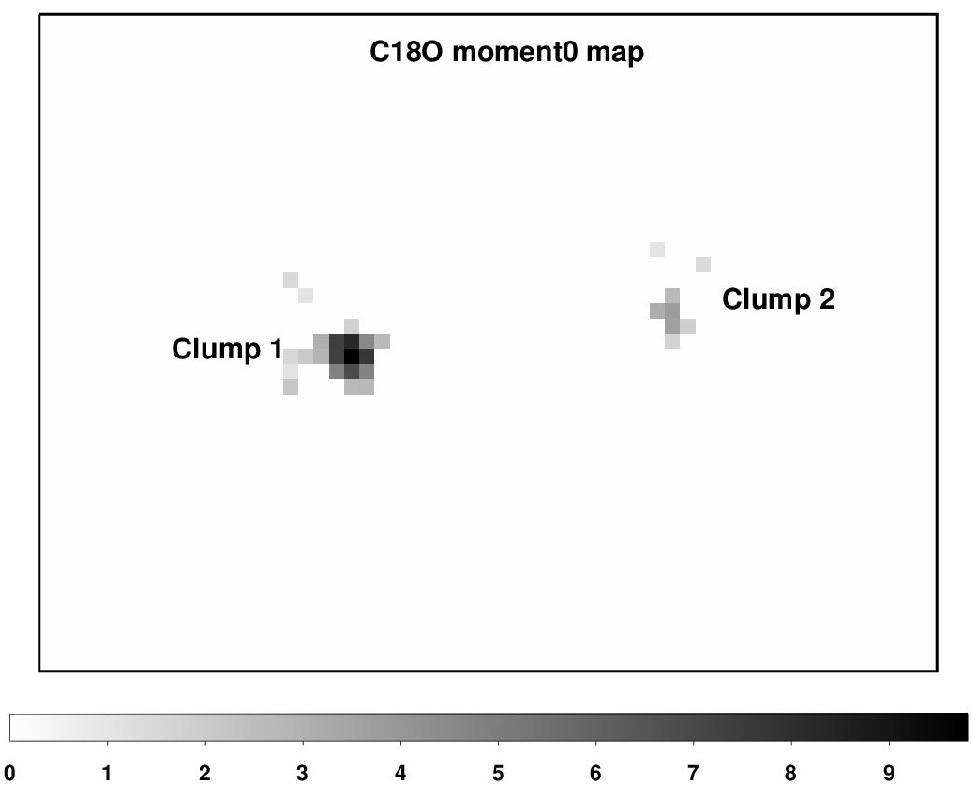}}
	\caption{Velocity integrated intensity maps of the S201 region showing the two clumps 
	(identified as clumps 1 and 2) in $^{13}$CO (left) and C$^{18}$O (right). 
The color scales correspond to the velocity-integrated intensity in K km s$^{-1}$. 
	The central coordinates of the maps are $\alpha_{J2000}$~$=$~03$^{h}$03$^{m}$12$\fs$72, 
	$\delta_{J2000}$~$=$~$+$60$\degr$28$\arcmin$08$\farcs$01 and the dimensions are $\sim7\arcmin\times\sim5\arcmin$.
	North is up and east is to the left.}\label{fig:moment0maps}
\end{figure*}

The $^{13}$CO gas traces the extended low-density gas around the clumps, whereas 
C$^{18}$O traces the highly compact central dense regions of the clumps (see Figure \ref{fig:moment0maps}). 
To demonstrate this, we estimate the optical depths, column densities of $^{13}$CO and C$^{18}$O , 
and the resultant $H_{2}$ column densities in Appendix \ref{sec:tauco_appendix}; these values are listed in Table \ref{tab:cotaucold}. 
The optical depths suggest that 
C$^{18}$O is optically thin and hence traces the densest parts of the clumps, 
thereby revealing the level of turbulence at the densities traced by 
850 $\mu$m dust emission. 

We further derive the one-dimensional thermal velocity dispersion ($\sigma_{\mathrm{T}}$) implied by 
the kinetic temperature of the C$^{18}$O gas using the 
relation
\begin{equation}
\sigma_{T} = \sqrt{\frac{k\,T_{\mathrm{kin}}}{{M_{\mathrm{C^{18}O}}}}},
\end{equation} 
where $k$ is the Boltzmann constant, $T_{\mathrm{kin}}$ is gas kinetic temperature, which is equivalent to 
the mean dust temperatures ($T_{\mathrm{d}}$~$=$~27 K and 29 K for clump 1 and 2; Table \ref{tab:paramscl12}) 
under the assumption that the gas and dust are in local thermal equilibrium. $M_{\mathrm{C^{18}O}}$ is the mass of the C$^{18}$O molecule and is considered to be 30 amu. 
The estimated $\sigma_{\mathrm{T}}$ values are found to be 0.087$\pm$0.024 km s$^{-1}$ and 0.090$\pm$0.017 km s$^{-1}$ for clumps 1 and 2. 
Finally, the non-thermal velocity dispersion ($\sigma_{\mathrm{NT}}$), which is due to the turbulence, is estimated 
by correcting for thermal velocity dispersion using the relation
\begin{equation}
\sigma_{\mathrm{NT}} = \sqrt{ {\sigma_{\mathrm{V_{LSR}}}}^{2} - {\sigma_{\mathrm{T}}}^{2}}.
\end{equation}
The derived $\sigma_{\mathrm{NT}}$ values are 0.68$\pm$0.02 km s$^{-1}$ and 0.59$\pm$0.04 km s$^{-1}$ for clumps 1 and 2, respectively. 
Therefore, in the following analysis we used our derived C$^{18}$O non-thermal velocity dispersions, $\sigma_{\mathrm{NT}}$, to estimate 
the B-field strength and pressure, turbulent pressure, Alfv{\'e}nic velocity, Alfv{\'e}n Mach number, virial mass, etc.

\subsection{Magnetic field strength}\label{bfieldstrength}

In the subsections above, number densities and gas velocity dispersion values were extracted. 
Here, we derive the angular dispersion of the B-fields in order to estimate the B-field strength and other crucial parameters.

The plane-of-sky component of B-field strength ($B_{pos}$) can be estimated using 
the Davis-Chandrasekhar-Fermi method \citep[hereafter DCF method]{Davis1951,ChandrasekharFermi1953}, which is based on the assumption that turbulence-induced Alfv{\'e}n waves will distort B-field orientations. 
This method assumes that the following two conditions hold: (a) 
the ratio of turbulent (${\delta}B$) to large-scale ordered ($B_{o}$) B-field components is proportional to 
the ratio of one-dimensional non-thermal velocity dispersion ($\sigma_{v}$) to Alfv{\'e}nic velocity ($V_{A}$~$=$~$B_{o}/\sqrt{4\pi\rho}$; 
where $\rho$ is the mass density), i.e., $\delta B/B_{o} \sim \sigma_{v}/V_{A}$ \citep{Hildebrandetal2009}, and (b) 
 ${\delta} B/B_{o} \sim \sigma_{\theta}$, where $\sigma_{\theta}$ is the dispersion in the measured B-field orientation. 
The DCF method can be applied to sets of Gaussian-distributed polarization angles with 
angular dispersions less than 25$\degr$ \citep{Ostrikeretal2001}. However, B-fields in regions 
altered by H\,{\sc{ii}} regions or dragged by gravity will generally exhibit multiple components, and so have a 
significantly higher angular dispersions 
 \citep[e.g.,][]{Arthuretal2011,Chapmanetal2011,Santosetal2014,Eswaraiahetal2017,ChenZhiweietal2017}. 
 As shown in Figure \ref{fig:histpa}, the histograms of B-fields being impacted 
 by the H\,{\sc ii} region feedback exhibit either a widely spread distribution of 
 angles or multiple distributions with conspicuously separate peaks. 
In such regions alternative methods must be employed to extract the underlying dispersion in polarization angles 
caused by magnetized turbulence. Progress has recently been made towards the accurate estimation of 
${\delta}B/B_{o}$ through statistical analyses of polarization angles. 
These analyses include the ``structure function" \citep{Hildebrandetal2009} and 
the ``auto-correlation function" \citep{Houdeetal2009} of polarization angles.
These are referred to as modified DCF methods, and used to estimate the B-fields in such regions.
 
\begin{figure*}
\centering
\resizebox{8.75cm}{6.125cm}{\includegraphics{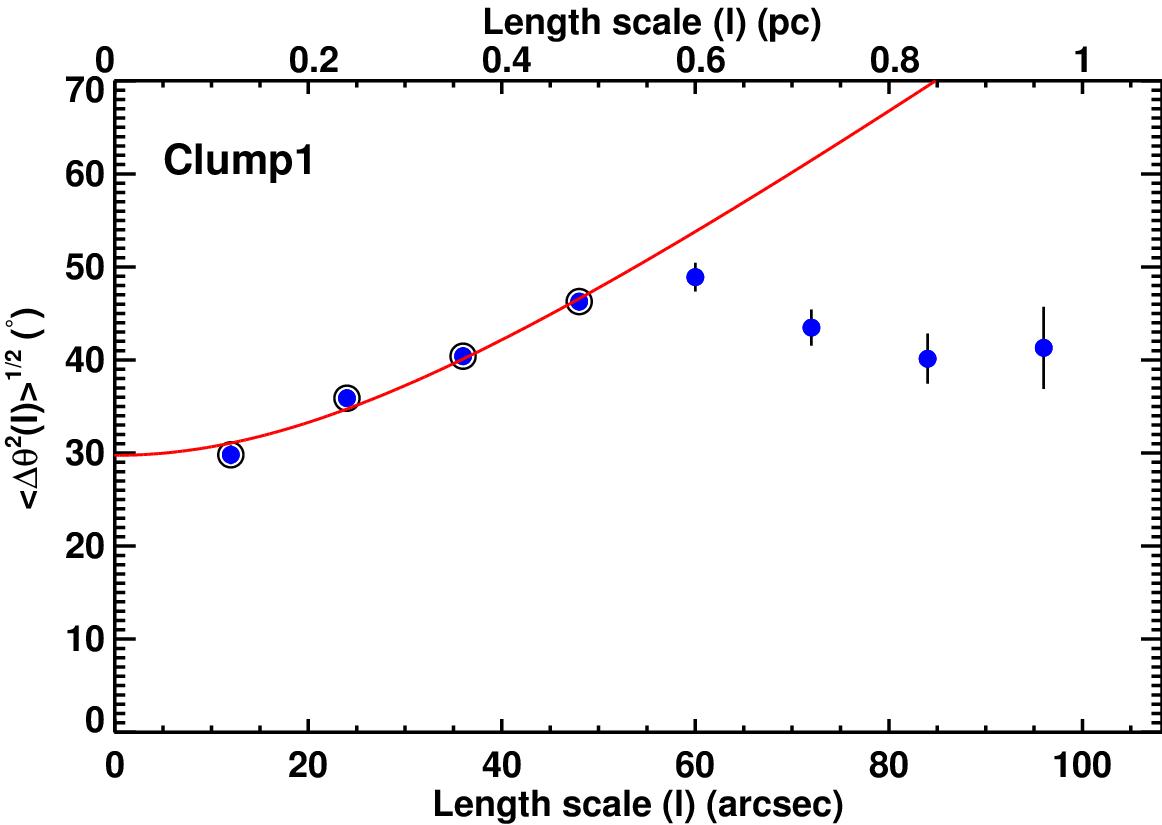}}
\resizebox{8.75cm}{6.125cm}{\includegraphics{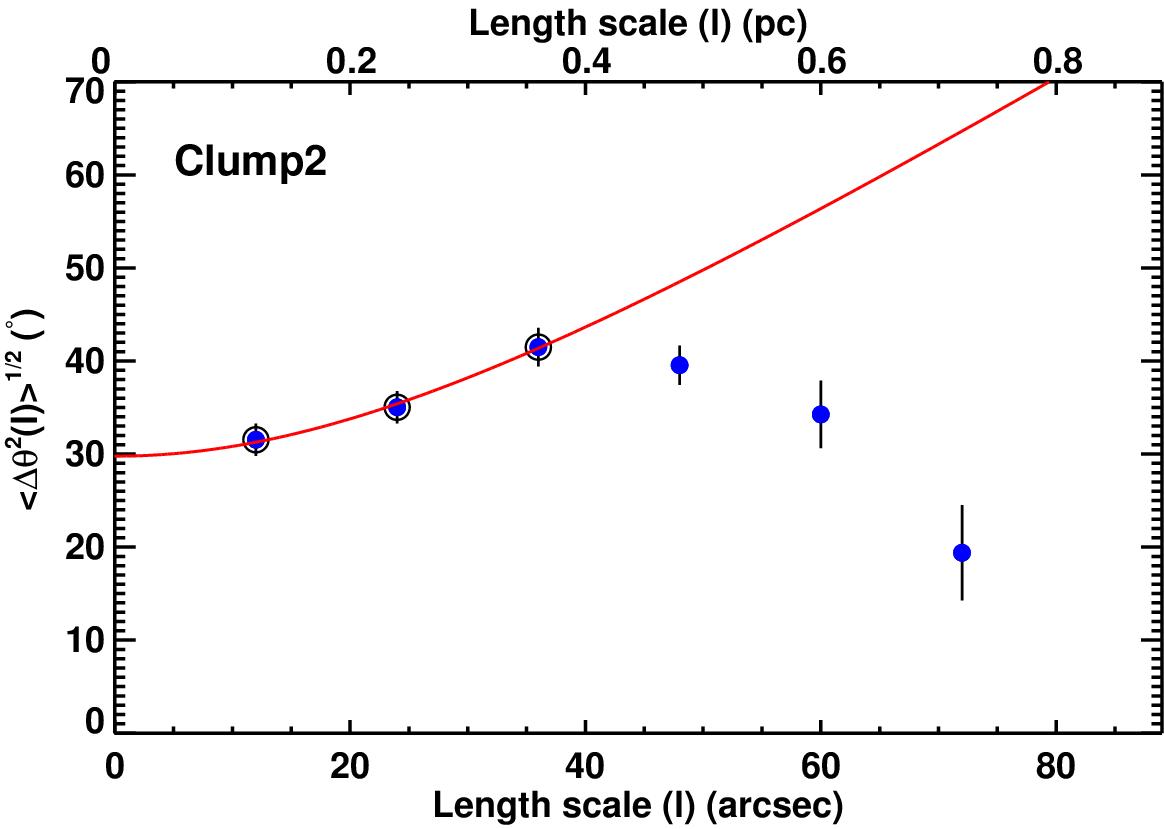}}
	\caption{Angular dispersion function versus length scale for clump 1 (left) and clump 2 (right). The plotted angular dispersions (blue filled circles) are 
	corrected for measured uncertainties. The best fits are shown with red lines. The data considered for the fits are depicted with encircled symbols.}\label{fig:sf}
\end{figure*}

\subsubsection{Structure function (SF) analysis}\label{subsec:sf_analyses}
In the structure function (SF) analysis \citep{Hildebrandetal2009}, the B-field is assumed to consist of a large-scale structured 
field, $B_o$, and a turbulent component, ${\delta}B$. The SF analysis demonstrates the variation of 
dispersion in position angles as a function of vector separation $l$. At some scale larger than the 
turbulent scale $\delta$,  ${\delta}B$ should reach its maximum value. At scales smaller than a scale $d$, 
the higher-order terms of the Taylor expansion of $B_0$ can be canceled out. 
When $\delta < l \ll d$, the SF follows the form: 
\begin{equation}\label{eq:adfrelation}
\langle \Delta \Phi ^2 (l)\rangle_{\mathrm{tot}} - \sigma_M^2(l) \simeq b^2 + m^2l^2.
\end{equation}
In this equation, $\langle \Delta \Phi ^2 (l)\rangle_{\mathrm{tot}}$, the square of the total measured 
dispersion function, consists of $b^2$, a constant turbulent contribution, $m^2l^2$, the contribution from 
the large-scale structured field, and $\sigma_M^2(l)$, the contribution of the measured uncertainty. 
The ratio of the turbulent to the large-scale component of the magnetic field is given by:
\begin{equation}\label{eq:adf_btbo}
\frac{\langle  {\delta}B^2 \rangle^{1/2}}{B_o} = \frac{b}{\sqrt{2-b^2}}.
\end{equation}
And $B_o$ is estimated using the modified DCF relation:
\begin{equation}\label{eq:adf_bfieldstrength}
B_o \simeq \sqrt{(2-b^2)4\pi \mu m_{\mathrm{H}} n_{\mathrm{H_2}}} \frac{\sigma_v}{b}.
\end{equation}
Then the estimated plane-of-sky magnetic field strength is corrected by a factor $Q$ 
\begin{equation}\label{eq:adf_qcorrection}
	B_{\mathrm{pos}} = Q~B_{0}
\end{equation}
where $Q$ is considered as 0.5 based on 
studies using synthetic polarization maps generated from numerically simulated clouds \citep{Ostrikeretal2001}, which 
suggest that B-field strength is uncertain by a factor of two when the B-field dispersion is $\leq$ 25$\degr$. 

The blue filled circles plotted in Figure \ref{fig:sf} represent 
the angular dispersions corrected for uncertainties 
($\langle \Delta \Phi ^2 (l)\rangle_{\mathrm{tot}} - \sigma_{M}^2(l)$) as a function of 
length scale, measured from the polarization data. The bin size of 12$\arcsec$, used in Figure \ref{fig:sf}, 
corresponds to the grid size of the polarization catalog yielded by {\it pol2map}. 
The maximum value in the current SF is lower than 
the value expected for a random field \citep[52$\degr$,][]{Poidevinetal2010}. 
Equation \ref{eq:adfrelation} is fitted to the data 
using the IDL {\sc MPFIT} nonlinear least-squares fitting algorithm \citep{Markwardt2009}. 
The resulting $\frac{\langle  {\delta}B^2 \rangle^{1/2}}{B_o}$ values are 0.40$\pm$0.02 and 0.40$\pm$0.07 for clumps 1 and 2, and 
the corresponding $B_{pos}$ strengths are derived using equations \ref{eq:adf_btbo}, \ref{eq:adf_bfieldstrength}, and \ref{eq:adf_qcorrection} to be 147$\pm$15 $\mu$G and 65$\pm$6 $\mu$G.

\begin{figure*}
\centering
\resizebox{8.75cm}{10.5cm}{\includegraphics{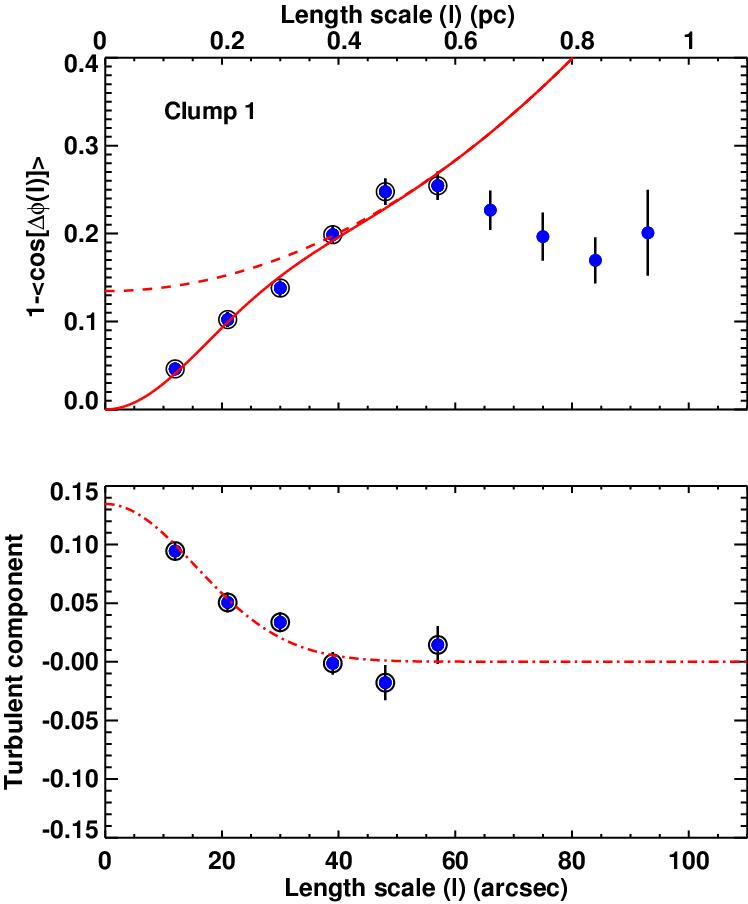}}
\resizebox{8.75cm}{10.5cm}{\includegraphics{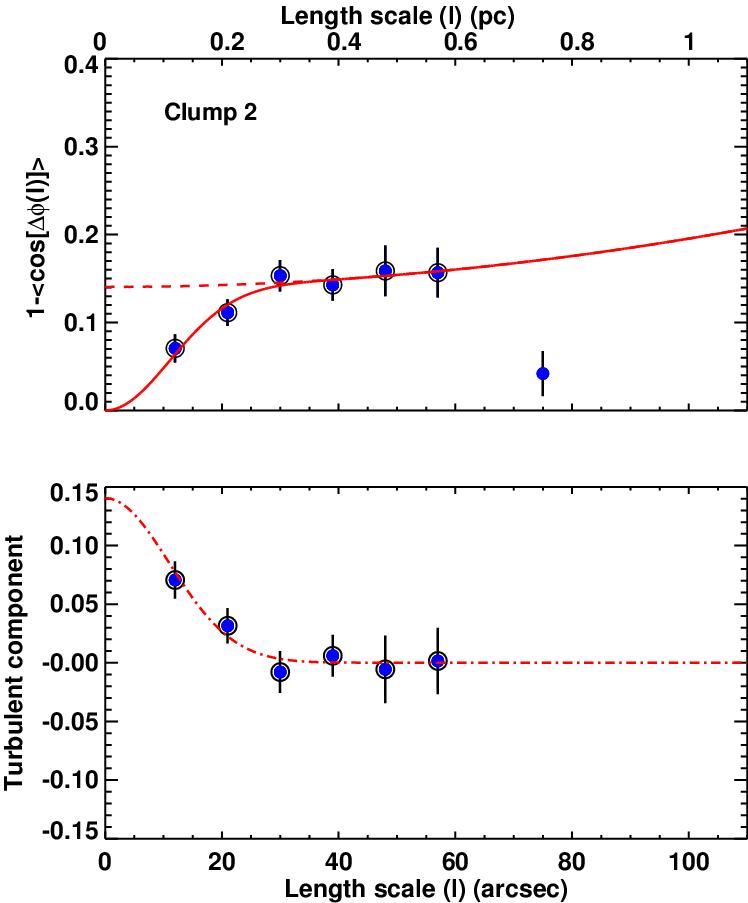}}
	\caption{ (Top) Auto-correlation function versus length scale for clump 1 (left) and clump 2 (right).
	Angular dispersions are shown as filled circles. 
	The red continuous line shows the best fit dispersion function and the red dashed line shows the 
	large-scale ordered or non-correlated components (1/N) ($\left<\delta B^{2}\right>/\left< B_{o}^{2}\right>$) + $a_{2}^{'} l^{2}$. 
	The data used in the fits are depicted with encircled filled symbols. (Bottom) Best fit turbulent or correlated component 
	(1/N) ($\left<\delta B^{2}\right>/\left< B_{o}^{2}\right>$) $e^{-l^{2}/2(\delta^{2}+2W^{2})}$ for clump 1 and clump 2, shown with dot-dashed lines. Plotted data points correspond to the difference between the derived angular dispersions 
	(encircled filled symbols; top panels) and the large-scale ordered component (red dashed line; top panels).} \label{fig:acf}
\end{figure*}

\subsubsection{Auto-correlation function (ACF) analysis}\label{subsec:acf_analyses}

Auto-correlation function (ACF) analysis \citep{Houdeetal2009} is an extension of SF analysis, 
which includes the effect of signal integration along the line of sight as well as within the beam. 
According to \citet{Houdeetal2009}, the ACF can be written as: 
\begin{equation}\label{eq:acf}
1 - \langle \cos \lbrack \Delta \Phi (l)\rbrack \rangle \simeq \frac{1}{N} \frac{\langle {\delta}B^2 \rangle}{\langle B_0^2 \rangle} \times \lbrack 1 - e^{-l^2/2(\delta^2+2W^2)}\rbrack + a_2\arcmin l^2,
\end{equation}
where $\Delta \Phi (l)$ is the difference in position angle of two vectors separated by 
a distance $l$, $W$ is the beam radius (6.0$\arcsec$ for the JCMT, i.e., the FWHM beam size of 14$\arcsec$ 
divided by $\sqrt{8 \ln{2}}$), $a_2\arcmin$ is the slope of the second-order term of 
the Taylor expansion, and $\delta$ is the turbulent correlation length. 
$N$ is the number of turbulent cells probed by the telescope beam and is given by:
\begin{equation}
N = \frac{(\delta^2 + 2W^2)\Delta\arcmin}{\sqrt{2\pi}\delta^3},
\end{equation}
where $\Delta\arcmin$ is the effective thickness of the cloud. The ordered magnetic field strength is given by 
\begin{equation}\label{eq:bfield_acf}
B_o \simeq \sqrt{4\pi \mu m_{\mathrm{H}} n_{\mathrm{H_2}}} \sigma_v \left[ \frac{\langle  {\delta}B^2 \rangle}{\langle B_o^2 \rangle} \right]^{-1/2}.
\end{equation}

The top panels of Figure \ref{fig:acf} show the angular dispersion functions of the polarization 
vectors in clumps 1 and 2, while the bottom panels show the respective correlated components of 
the dispersion functions. 
In our fitting, $\Delta\arcmin$ is set to 
$31\arcsec\pm1\arcsec$ for clump 1 and  
$36\arcsec\pm1\arcsec$ for clump 2 (these correspond to the effective thickness of the clumps, derived using the relation $\sqrt{\delta_{a} \times \delta_{b}}$; where $\delta_{a}$ and $\delta_{b}$ are the FWHMs of the major and minor axes, respectively (see Table \ref{tab:paramscl12}). 

Equation \ref{eq:acf} is fitted to the ACF data shown in Figure \ref{fig:acf} for clumps 1 (top panel) and 2 (bottom panel). 
The bin width for constructing the ACF ($1 - \langle \cos \lbrack \Delta \Phi (l)\rbrack \rangle$) function was chosen to be 9$\arcsec$. Various bin widths were investigated; we found that 
best fit was achieved with a bin width of 9$\arcsec$. 
The IDL {\sc MPFIT} non-linear least-square fitting algorithm \citep{Markwardt2009} was used and simultaneously constrained 
the three fitting parameters (i) $\delta$, (ii) $\frac{\langle  {\delta}B^2 \rangle}{\langle B_o^2 \rangle}$, and (iii) 
${a_{2}}\arcmin$; the best-fit values are listed in Table \ref{tab:paramscl12}. Using the fitted parameter 
$\frac{\langle  {\delta}B^2 \rangle}{\langle B_o^2 \rangle}$, along with derived parameters such as number densities and 
velocity dispersions, we have estimated the B-field strength using the modified DCF relation (equation \ref{eq:bfield_acf}). 
The estimated B-field strengths are 266$\pm$32\,$\mu$G and 
61$\pm$31\,$\mu$G for clumps 1 and 2 respectively. 
The turbulent correlation length $\delta$ is $13\pm3\arcsec$ (126$\pm$29 mpc) and $7\pm4\arcsec$ (68$\pm$36 mpc) for clumps 1 and 2.
The number of turbulent cells (N) is derived to be 1.4$\pm$0.4 and 5.0$\pm$4.8 for clumps 1 and 2, respectively. The derived parameters are listed in Table \ref{tab:paramscl12}.

In summary, the SF and ACF analyses yielded two B-field strengths: 
for clump 1 these differ by a factor of $\sim$2 (147$\pm$15 $\mu$G and 266$\pm$32 $\mu$G), 
while they are nearly identical for clump 2 (65$\pm$6 $\mu$G and 61$\pm$31 $\mu$G),
although the ACF-derived B-field strength has $\sim$50\% uncertainty. 
It should be noted here that the measured uncertainties in the B-fields strengths are 15 $\mu$G and 32 $\mu$G for clump 1 and 6 $\mu$G and 31 $\mu$G for clump 2. These are random errors, resulting from 
propagation of the uncertainties of various parameters (such as gas number density, 
gas velocity dispersion, and B-field angular dispersion) through the modified DCF formulae. We caution here that, 
in practice these measurement errors can be dominated by systematic errors associated with 
various parameters such as dust opacity, dust temperature, flux calibration, 
clump geometry (and hence depth), and inclination angle, etc. 
\citep[e.g.,][]{Pattleetal2017b,WangJWetal2019,LiuJunhaoetal2019}. These factors could be more severe in clump 1 because of the existence of more extended emission around the peak of dust emission. Therefore the real uncertainties could be much larger than the quoted errors in B-field strengths.

It is clear from Sections \ref{subsec:acf_analyses} and \ref{subsec:sf_analyses} above (see also Table \ref{tab:paramscl12}) that for clump 2 the ACF-yielded parameters 
have higher uncertainties in comparison to those yielded by SF. This could be attributed to the relatively small  number of vectors in clump 2 causing ACF to fail constraining all the fitting parameters simultaneously (see column 3 of Table \ref{tab:paramscl12} for ACF). In addition, since the column density is relatively lower, the beam dilution and signal integration effects may not be important in clump 2. Conversely, for clump 1 these factors seem to be crucial and taken care of ACF. Therefore, we have used B-field strengths derived from ACF (266$\pm$32 $\mu$G) for clump 1 and from SF (65$\pm$6 $\mu$G) for clump 2 in the further analyses.

\subsection{Ionized gas properties: thermal and radiation pressure in S201}\label{subsec:radio_ptherm}

\begin{figure}
{\includegraphics[width=8.5cm]{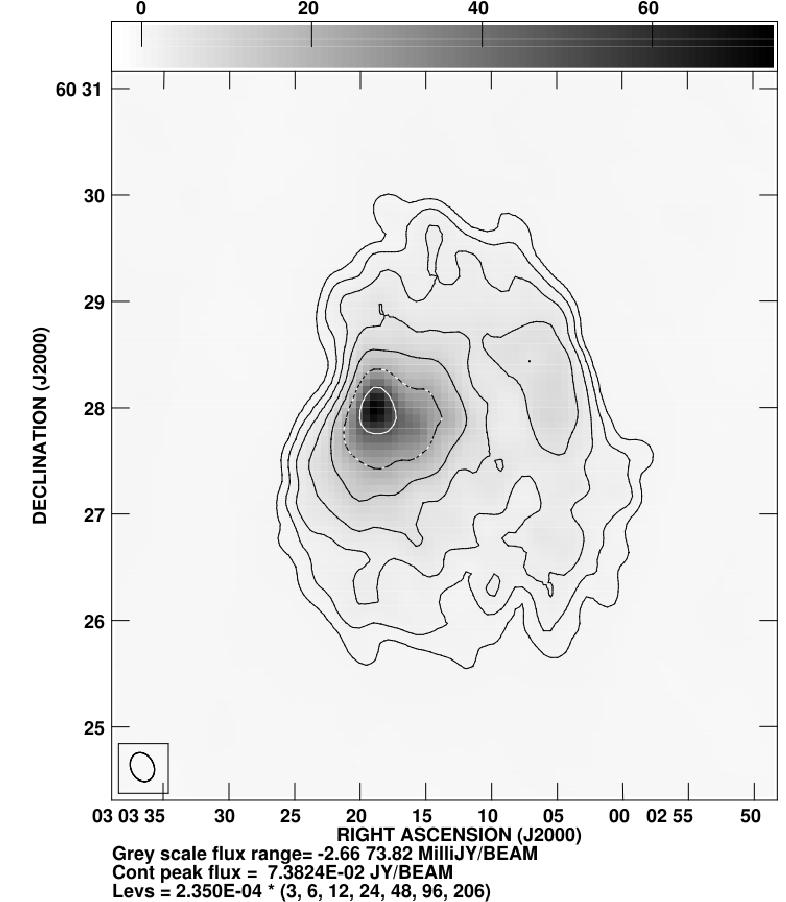}}
 \caption{A radio continuum map of S201 at 1.4 GHz.
	The contour levels are at $[$3, 6, 12, 23, 48, 96, and 202$]$ $\times$ 2.3$\times$10$^{-4}$ Jy/beam, 
	where 2.3$\times$10$^{-4}$ Jy/beam ~is the rms noise of the map. 
	The VLA beam size is shown in the lower-left corner of the figure, and is $\sim17\arcsec \times \sim13\arcsec$.}
	\label{fig:radioconti}
\end{figure}

Figure \ref{fig:radioconti} shows the radio continuum view of the S201 H\,{\sc{ii}} region 
at VLA 21~cm/1.4 GHz. The flux density (S$_{\nu}$) of S201, estimated by integrating over 
3$\sigma$ contours, is $\sim$1.0$\pm$0.1 Jy, where $\sigma$ is the rms noise of the 21~cm map. 
Our 21~cm flux density is, within uncertainty, consistent with flux density at 6~cm 
(1.2$\pm$0.2 Jy; \citealt{Fellietal1987}, and references therein). 
The flux densities at 21~cm and 6~cm reflect a flat spectrum, indicating that the nebula is optically 
thin at 21~cm. Considering 8302 K as the electron temperature ($T_{\mathrm e}$; \citealt{Balseretal2011}) 
and $\sim$120$\arcsec$ (or 1.2~pc at 2~kpc) 
as the effective radius of the ionized gas (Figure \ref{fig:radioconti}), 
we estimated the average electron density ($n_{e}$) of the S201 H\,{\sc{ii}} 
region using the following equations \citep{Martin-Hernandezetal2005}
\begin{scriptsize}
\begin{equation}\label{eq_flux}
n_{e}  = \frac{4.092\times 10^{5}\mathrm{ cm}^{-3}}{\sqrt{b(\nu,T_e)}}
\left(\frac{S_\nu}{\mathrm{Jy}}\right)^{0.5}\left(\frac{T_e}{10^4\mathrm{
    K}}\right)^{0.25}\left(\frac{D}{\mathrm{kpc}}\right)^{-0.5}\left(\frac{\theta_D}{\arcsec}\right)^{-1.5}
\end{equation}
\end{scriptsize}

and

\begin{scriptsize}
\begin{equation}
b(\nu,T_e) = 1+0.3195\log\left(\frac{T_e}{10^4\mathrm{K}}\right)-0.2130\log\left(\frac{\nu}{\mathrm{GHz}}\right), 
\end{equation}
\end{scriptsize}
where $S_\nu$ is the radio continuum integrated flux at frequency
$\nu$, $\theta_{D}$ is the angular diameter of the source, $D$ (2 kpc) is the
distance from the Sun, and $T_{e}$ is the electron temperature in the
ionized plasma. Using this formalism, we estimated $n_{e}$ as 226$\pm$11~cm$^{-3}$.  
We then estimated the corresponding thermal pressure, due to ionized gas distributed over the area of diameter $\sim$120$\arcsec$, to be (5.2$\pm$0.3)~$\times$~10$^{-10}$~dyn cm$^{-2}$ 
using the relation $P_{\mathrm{te}} = 2 n_{e} k_{b} T_{e}$. 

The mean thermal pressure given above my be valid for the entire region surrounded by the H\,{\sc ii} emission. The radio emission is observed to be uneven in the region of S201 (cf. Figure \ref{fig:radioconti} 
and also see Figure \ref{fig:diffpaBIG}) in the sense that more H\,{\sc ii} emission is concentrated close to clump 1 than is concentrated near clump 2. Because of this, the relative impact of H\,{\sc{ii}} emissions, in terms of the thermal pressures acting on the 
clump surfaces, will be different. Therefore, we estimate average thermal pressures close to the clumps. 
The integrated fluxes, $S_\nu$, within the green contours (see Figure \ref{fig:diffpaBIG}) 
extended over circular diameters of $\sim$66$\arcsec$ and $\sim$48$\arcsec$, 
are found to be 0.4$\pm$0.2\,Jy and 0.05$\pm$0.01\,Jy for clumps 1 and 2, respectively. 
Using these parameters, along with the value of ~$T_{e}$ quoted above, 
we estimated the electron densities, $n_{e}$, 
as 360$\pm$80\,cm$^{-3}$ and 207$\pm$14\,cm$^{-3}$, and the 
resultant thermal pressures, $P_{te}$, to be (8$\pm$2)~$\times$~10$^{-10}$~dyn cm$^{-2}$ and 
(4.7$\pm$0.3)~$\times$~10$^{-10}$~dyn cm$^{-2}$ for clumps 1 and 2, respectively. 

We estimated the mean radiation pressure ($P_{rad}$)
driven by an ionizing star of spectral type O6V \citep{Ojhaetal2004,Deharvengetal2012}. 
The ionizing flux emitted by one~O6V star is $q_{0}$~$=$~4.15$\times10^{10}$~photons~cm$^{-2}$~s$^{-1}$ 
\citep{Sternbergetal2003}, and each UV photon carries an energy h$\nu$~$=$~20~eV,
so the estimated $P_{rad}$, using the relation $P_{rad}$~$=$~$h\nu~q_{0}/c$,
is $0.44\times10^{-10}$~dyn~cm$^{-2}$.
We note here that the same $P_{rad}$ value is used for both clumps and in both cases is negligible
in comparison to the thermal pressures. 
The derived thermal and radiation pressure values are listed in Table \ref{tab:paramscl12}. 

\section{Discussion}\label{sec:discuss}

Here we discuss the interplay amongst various key parameters, clump stability 
based on virial and critical mass estimations, and their 
relevance to the formation and evolution of clumps as well as to the feedback process. 

\subsection{B-fields versus Turbulence}\label{bpturbp}

For clump 1 the magnetic and turbulent pressures, estimated using the relations 
$P_{B}$~$=$~$B^{2}/8\pi$ and $P_{\mathrm{turb}}$~$=$~$\rho{\sigma_{\mathrm{NT}}}^{2}$ (c.f. Section \ref{subsec:gasveldisp}), 
are found to be (28\,$\pm$\,7)$\times$10$^{-10}$ dyn~cm$^{-2}$ and
(11\,$\pm$\,2)$\times$10$^{-10}$ dyn~cm$^{-2}$, respectively. 
Similarly, for clump 2, these values are found to be 
(1.7\,$\pm$\,0.3)$\times$10$^{-10}$ dyn~cm$^{-2}$ and (2.1\,$\pm$\,0.4)$\times$10$^{-10}$ dyn~cm$^{-2}$.
The magnetic to turbulent pressure ratios, $P_{\mathrm{B}}$/$P_{\mathrm{turb}}$, are
estimated to be 2.6\,$\pm$\,0.8 and 0.8\,$\pm$\,0.2 for clump 1 and 2, respectively. This suggests that B-fields dominate over turbulence in clump 1, whereas turbulence dominates over B-fields in clump 2. 

The turbulent Alfv{\'e}n Mach number ($M_{A}$) describes the 
relative importance of B-fields compared to that of turbulence, and
hence it is a key parameter in models of cloud formation and evolution 
\citep[e.g.,][]{Ostrikeretal2001,Padoanetal2001,NakamuraLi2008}. 
In the sub-Alfv{\'e}nic case ($M_{A} \leqslant$ 1), B-fields are strong enough
to regulate turbulence, causing an organized B-field orientation. 
In the super-Alfv{\'e}nic case ($M_{A} >$ 1), 
the turbulence is capable of perturbing the morphology of B-fields.

The Alfv{\'e}nic velocity, $V_{\mathrm{A}}$~$=$~$\frac{B_{\mathrm{los}}}{\sqrt{4\pi\,\rho}}$ (where $\rho$~$=$~$n_{H_{\mathrm{2}}}\,\mu\,m_{\mathrm{H}}$),
is estimated to be 1.6$\pm$0.2 km s$^{-1}$ for clump 1 and 0.7$\pm$0.1 km s$^{-1}$ for clump 2.
The Alfv{\'e}n Mach number, $M_{\mathrm{A}}$~$=$~$\sqrt{3} (\frac{\sigma_{\mathrm{NT}}}{V_{\mathrm{A}}})$, 
is estimated to be 0.8$\pm$0.1 for clump 1 and 1.4$\pm$0.2 for clump 2. 
These estimates agree with the findings above based on the ratios of $P_{\mathrm{B}}$/$P_{\mathrm{turb}}$, and imply that turbulent motions are sub-Alfv{\'e}nic and super-Alfv{\'e}nic in clumps 1 and 2, respectively.  

\subsection{B-field versus Thermal pressure by H\,{\sc ii} region}\label{subsec:bpthermp}

The ratio of magnetic (c.f. Section \ref{bpturbp}) to thermal pressures (c.f. Section \ref{subsec:radio_ptherm}), 
P$_{\mathrm{B}}$/P$_{\mathrm{te}}$, is found to be 3$\pm$1 and 0.4$\pm$0.1 in clumps 1 and 2, respectively. 
These results imply two contrasting scenarios, in that 
B-fields dominate over thermal pressure acting on clump 1, whereas 
thermal pressure dominates over B-fields in clump 2. 
Evidently, B-fields are strong enough to control the expanding I-front in clump 1 and, conversely, 
the I-front is strong enough to dictate the B-fields in clump 2. 
The consequences of this interplay between B-fields and thermal pressures will be discussed in Section \ref{subsec:overallmorpcorrel}. 

\subsection{Clump stability: virial and critical mass ratios}\label{subsec:clumpstab_virial_critical_mass_ratios}

\subsubsection{Are the clumps gravitationally bound?}\label{subsubsec:virialanalysis}

Considering the case in which clumps are not supported by B-fields and also 
not confined by external pressure (in the form of envelope material around the clumps), 
we estimate the viral masses and virial mass ratios. 

In order for self-gravitating clumps to be in virial equilibrium, 
the following relation between gravitational potential energy
($|{\mathcal{G}}|$) and internal kinetic energy (${\mathcal{E}}$) should hold \citep{McKeeZweibel1992}

\begin{equation}
{\mathcal{G}} = 2 {\mathcal{E}},
\end{equation}
where ${\mathcal{E}}$~$=$~3/2 M$\sigma^{2}$.

The gravitational potential energy can be written as 
\begin{equation}\label{eq:gravpoten}
{\mathcal{G}} = -\frac{3}{5}\alpha\beta \frac{G M^{2}}{r},
\end{equation}

where $r$~$=$~$R_{\mathrm{eff}}$ and $G$ is the gravitational constant.
$\alpha$ corresponds to a geometric factor, a function of eccentricity, 
and $\beta$ is a function of the power-law index of the density profile
($\rho$~$\propto$~r$^{-a}$, where $a$~$=$~1.6 for an isothermal cloud in equilibrium; \citealt{Bonnor1956}). 
More details on deriving these factors, assuming that the clumps are prolate ellipsoids, can be
found in \citet[][and references therein]{LiDietal2013}. For our given values of 
$\sigma$ and $r$, using the above equations,
the virial mass ($M_{vir}$) can be estimated using the relation
\begin{equation}\label{eq:virialmass}
	M_{vir} = \frac{5}{\alpha \beta} \frac{\sigma^{2} r}{G}.
\end{equation}
Taking $\sigma$~$=$~$\sigma_{\mathrm{V_{LSR}}}$ of C$^{18}$O and $r$~$=$~$R_{\mathrm{eff}}$, $M_{vir}$ is estimated to be 50 $M_{\sun}$ and 40 $M_{\sun}$ for clump 1 and clump 2, respectively.
For our estimated masses ($M$; 191 $M_{\sun}$ for clump 1 and 30 $M_{\sun}$ for clump 2; 
c.f. Section \ref{subsec:cold_numb_mass}), 
the derived virial mass ratios, $R_{\mathrm{vir}}$~$=$~$M$/$M_{\mathrm{vir}}$ 
are 3.9 and 0.7 for clumps 1 and 2. 
Therefore, clump 1 is bound by gravity and may collapse once it becomes unstable, whereas clump 2 is gravitationally 
unbound. 

\subsubsection{Stability and Critical Mass based on Turbulence, Temperature, and B-fields}\label{subsubsec:stability_and_critical_mass}

Critical mass $M_{C}$ is the maximum mass that can be supported by the combined contributions of internal velocity dispersion 
(contribution from turbulence and neutral gas temperature, i.e., non-thermal and thermal contributions) 
and B-field in the clump. The two effects can be represented as 

\begin{equation}
	M_{C} = M_{J} + M_{\mathrm{\phi}}, 
\end{equation}
which is accurate within 5\% to results from more rigorous calculations \citep{McKee1989}. 

The Jean mass for a non-magnetic isothermal cloud \citep{Bonnor1956,McKeeZweibel1992} is 
\begin{equation}\label{eq:jeans_mass}
	M_{J} = 1.182 \frac{{C_{\mathrm{eff}}}^{4}}{G^{3/2}{P_{\mathrm{env}}}^{1/2}}, 
\end{equation}
where $C_{\mathrm{eff}}$~$=$~$\sqrt{{C_{\mathrm{s}}}^2+{\sigma_{\mathrm{NT}}}^2}$. 
The thermal sound speed $C_{\mathrm{s}}$~$=$~$\sqrt{\frac{k\,T_{\mathrm{kin}}}{\mu_{\mathrm{H}}\,m_{\mathrm{H}}}}$ 
was estimated to be 0.28$\pm$0.01 km s$^{-1}$ for clump 1 and 0.29$\pm$0.01 km s$^{-1}$ for clump 2. In this equation,
$T_{\mathrm {kin}}$ = $T_{\mathrm {dust}}$ and is 27 K for clump 1 and 29 K for clump 2. 
The $C_{\mathrm{eff}}$ values are estimated as 0.74$\pm$0.02 km s$^{-1}$ and 0.66$\pm$0.04 km s$^{-1}$ for clumps 1 and 2. 

In Equation \ref{eq:jeans_mass}, the envelope pressure (or external pressure) caused by the 
low-density $^{13}$CO gas can be estimated as 
\begin{equation}
	P_{\mathrm{env}} =  n_{\mathrm{env}}\,\mu\,m_{\mathrm{H}}\,{\sigma_{\mathrm{env}}}^2, 
\end{equation}
where $\sigma_{\mathrm{env}}$ is the velocity dispersion in the low-density envelope, 
which we treated as $\sigma_{\mathrm{env}}$~$=$~$\sigma_{\mathrm{V_{LSR}}}$ of the $^{13}$CO gas 
(c.f. Table \ref{tab:gaussfitresults}), and the corresponding mean gas number densities 
$n_{\mathrm{env}}$ were estimated over larger extents to be 
$\sim$0.9$\times$10$^{4}$ cm$^{-3}$ and $\sim$0.5$\times$10$^{4}$ cm$^{-3}$ for clumps 1 and 2. 

The maximum mass that can be supported by B-fields will be 
\begin{equation}
	M_{\mathrm{\phi}} = c_{\mathrm{\phi}} \frac{\pi B r^{2}}{G^{1/2}},
\end{equation}
where $c_{\mathrm{\phi}} \sim 0.12$ according to numerical simulations \citep{Tomisakaetal1988}.

The estimated critical masses $M_{C}$ are 78 $M_{\sun}$ and 48 $M_{\sun}$ for clumps 1 and 2. 
The critical mass ratios $R_{C}$~$=$~$M/M_{C}$ are found to be 2.5 and 0.6 for clumps 1 and 2. 
These results suggest that for clump 1 the support rendered by the combined contributions from thermal gas energy, turbulence, and B-fields is not sufficient 
to counteract gravity, whereas the opposite situation prevails in clump 2, 
such that its stability is determined by thermal energy, turbulence, and 
B-fields rather than by self-gravity. This picture is further corroborated by the presence of a larger number 
of Class 0 and I source in and around clump 1. In contrast, clump 2 is inactive as 
there exist no YSOs at its center, except for a few Class I sources formed at its boundary 
(see Figure \ref{fig:tddustysos}). 

\subsection{Compressed B-fields and enhanced B-field strength in the clumps}\label{subsec:enhancedB}

The B-field strength in clump 1 (266$\pm$32 $\mu$G) is larger by a factor of $\sim$4 than that in clump 2 (65$\pm$6 $\mu$G).  
Additionally, the spread in ADF values ($\sim$30~--~$\sim50\degr$;~Figure \ref{fig:sf}) as well as ACF values ($\sim$0.05~--~$\sim$0.37; top panels of 
Figure \ref{fig:acf}) for clump 1 are relatively higher than those of clump 2 (ADF range $\sim$30~--~$\sim$40$\degr$; 
while ACF range $\sim$0.03~--~$\sim$0.16). Furthermore, 
the spread in the offset angles ($\Delta(\theta)$) between B-fields ($\theta_{B}$) and intensity gradients ($\theta_{IG}$) in clump 1 is relatively larger than that of clump 2. These signatures imply that B-fields are more curved and {\it draped} around clump 1, thereby following 
a bow-like structure (see Figure \ref{fig:polvecmaps}). Similar features have been seen in clump 2, 
but with a lesser degree of curvature. \citet{Aluzasetal2014}, based on 2D MHD simulations, show that when an oblique shock
interacts with an isolated cylindrical cloud, the B-fields wrap around the cloud, attaining a roughly circular shape (see their Fig. 1b) similar to the B-field morphologies observed in clump 1 (Figure \ref{fig:polvecmaps}).
Based on 3D radiation-magnetohydrodynamic (RMHD) simulations of
pillars and globules, \citet{MackeyLim2011b,Mackey2012,MackeyLim2013} have shown that in the case of initially strong B-fields (160 $\mu$G) oriented perpendicular to an I-front, field lines
at the head of a cometary globule get compressed into a curved morphology, by closely following the
bright rim in a similar manner to our present observations towards clumps 1 and 2 (see Figure \ref{fig:polvecmaps}). 
These findings are consistent with other observations and MHD simulations 
\citep[e.g.,][]{Lyutikov2006,DursiPfrommer2008,PfrommerJonathanDrusi2010,Arthuretal2011,Santosetal2014,Kusuneetal2015,Klassenetal2017}.

As can be seen from Figure \ref{fig:diffpaBIG}(a), contours of dust and ionized emission are closely spaced in 
their zone of interaction, suggesting possible compression in 
the interface between the hot and cold media. 
This interaction has also compressed B-fields and enhanced their strength. 
Consequently, the stronger B-fields may shield the clumps from the H\,{\sc ii} region. 
In the perpendicular field case, the B-fields get amplified directly upstream of the cloud
where the flow stagnates against it, where B-field pressure and field tension
continue to build \citep[][see their Fig. 1c]{Aluzasetal2014}. 
In MHD simulations, when B-fields get compressed their strengths become enhanced
in the shells or clumps by a factor of about 5 to 6 compared to those inside
the expanding H\,{\sc{ii}} regions \citep[e.g.,][see also \citealt{Wareingetal2017} for a similar enhancement
of B-field strength in environments with mechanical stellar feedback]{MacLowetal1994,Gregorietal1999,Klassenetal2017}.
Therefore, our results provide evidence for enhanced B-field strengths in clumps, 
which we attribute to the effect of thermal pressure. 
This is because the more the H{\sc ii} region interacts with the cloud, the more the field lines are compressed, 
resulting in a considerable enhancement in B-field strengths. 

Higher values of B-field strength around $\sim$50~--~$\sim$400~$\mu$G, have 
been measured at the edges of H\,{\sc{ii}} regions using H{\sc I}/OH 
Zeeman measurements \citep{Trolandetal1986,Trolandetal2016,MayoTroland2012} 
as well as dust extinction \citep[e.g.,][]{Kusuneetal2015,Eswaraiahetal2017,ChenZhiweietal2017}, and emission polarization \citep[e.g.,][]{Valleeetal2005,Pattleetal2018}. 
For example, \citet{MayoTroland2012} have measured B-field strength of $\sim$80 $\mu$G using the H{\sc I} 
Zeeman effect in the photodissociation region (PDR\footnote{The PDR is
a thin layer lying between the molecular cloud and the H\,{\sc{ii}} region.}) 
DR 22, which they interpret as resulting from B-fields amplified by 
compression of the PDR due to absorption of the momentum of stellar radiation 
(also see \citealt{Pellegrinietal2007}, for a similar explanation).
Based on NIR polarimetry towards the bright-rimmed cloud SFO 74, \citet{Kusuneetal2015} 
have measured an enhanced B-field strength of $\sim$90 $\mu$G inside the tip 
rim due to UV-radiation-induced shock. 
Similarly, enhanced B-field strengths of 100~--~300~$\mu$G have been inferred in PDRs around ionized regions using radio recombination lines (RRLs; \citealt{Balseretal2016}). 
SCUBAPOL2 observations towards M16 show that the derived B-field strength lies between 170 $\mu$G and 320~$\mu$G \citep{Pattleetal2018}. 
Based on SCUBAPOL observations towards S106, \citet{Valleeetal2005} estimated that its B-field strengths 
lie between 240 $\mu$G and 1040 $\mu$G. 
Similarly, \citet{Robertsetal1995} conducted OH Zeeman measurements towards S106; 
their derived B-field strengths range from 100 $\mu$G to 400 $\mu$G. 
Evidently, our derived B-field strengths ($\sim$50~--~$\sim$200\,$\mu$G) towards the 
clumps in S201 are in close agreement with the values quoted in the literature. 

\subsection{The role of H\,{\sc{ii}} region feedback and its relation to the observed turbulence in the clumps}\label{subsec:bpturbp}

We compare the magnetic (P$_{\mathrm{B}}$) and turbulent (P$_{\mathrm{turb}}$) pressures of 
the clumps with the thermal pressure (P$_{\mathrm{te}}$) exerted on them by the H\,{\sc ii} region, 
and examine whether turbulence is being injected into the clumps by the H\,{\sc ii} region in the presence of 
B-fields of varying strengths. The comparison among these pressures suggests two relations: 
(i) P$_{\mathrm{B}}$~$>$~P$_{\mathrm{turb}}$~$>$~P$_{\mathrm{te}}$ in clump 1 and (ii) 
P$_{\mathrm{te}}$~$>$~P$_{\mathrm{turb}}$~$>$~P$_{\mathrm{B}}$ in clump 2. 
This implies the dominance of B-fields in clump 1 and of thermal pressure on clump 2. 

The stronger B-fields in clump 1 could be able to guide the I-front to escape from 
the filament-ridge and also shield the I-front from entering clump 1; 
as a consequence the shock strength will be reduced \citep[e.g.,][]{Krumholzetal2007}. 
Two-dimensional numerical simulations of the interactions 
between magnetized shocks and radiative clouds show that B-fields external to, but 
concentrated near, the surface of the cloud suppress the growth of destructive 
hydrodynamic instabilities \citep{Chandrasekhar1961,MacLowetal1994,Jonesetal1996,Fragileetal2005}, thereby shielding the cloud from 
erosion or destruction. Conversely, non-magnetized, nonradiative clouds are 
destroyed on a few dynamical timescales through hydrodynamic Kelvin-Helmholtz, Richtmyer-Meshkov, 
and Rayleigh-Taylor instabilities \citep[e.g.,][]{Kleinetal1994,Nakamuraetal2006}. 
Eventually, the B-field dominates over turbulence as is seen in clump 1 (c.f. Section \ref{bpturbp}). 
In contrast, due to the limited impact of the H\,{\sc ii} region on clump 2, the B-field strength has not been 
enhanced to higher values in this region. Therefore, we hypothesize that due to these relatively 
weak B-fields the expanding I-front might have driven a shock front into clump 2, resulting in a 
higher turbulence pressure compared to magnetic pressure, as is seen (c.f. Section \ref{bpturbp}). 

\subsection{Pressure balance between clumps and stellar feedback, and the consequences}\label{subsec:overallmorpcorrel}

Assuming that the primordial filament within which S201 and W5E complexes 
formed \citep{Deharvengetal2012}, followed a Plummer-like column density profile \citep{Arzoumanianetal2011,Juvelaetal2012,Palmeirimetal2013,Andreetal2019} 
and also that the primordial B-fields were threaded perpendicular to the filament's long axis 
\citep[e.g.,][]{Chapmanetal2011,Sugitanietal2011,LiHuabaietal2013,Wangetal2017,PlanckColloborationXXXVetal2016}, 
we discuss below the formation of clumps, enhanced gas and magnetic pressures, 
the pressure balance between clumps and feedback, and their consequences 
for the evolution of clumps and star formation in them, and the formation of bipolar H\,{\sc ii} regions. 

The expanding I-front from a deeply embedded H\,{\sc{ii}} region in a filament becomes anisotropic, such that 
the flows along the dense filament-ridge will become sonic as they are obstructed, 
while they are supersonic in the low-density regions both below and above the ridge. 
As a result of the natural anisotropic distribution of material in the filament, the H\,{\sc{ii}} region 
is lead to form bipolar bubbles \citep{Bodenheimeretal1979,FukudaHanawa2000}. 
Moreover, inclusion of B-fields in the filament would introduce 
an additional anisotropic pressure \citep{Tomisaka1992,Gaensler1998,PavelClemens2012,vanMarleetal2015}, 
because ionized material flowing along the B-fields will be accelerated, 
while those flows in the direction perpendicular to B-fields will be hindered due to the Lorenz force.
\citet{Krumholzetal2007}, based on MHD simulations, have shown that B-fields
suppress the sweeping up of gas perpendicular to field lines. 
As the H\,{\sc{ii}} region expands further into the cloud, gas and dust in the filament-ridge will be swept up, and as a result 
the accumulated material is led to form the dense clumps at the waist of H\,{\sc ii} region. 
It should be noted here that B-field strength will also be continuously enhanced due to flux freezing, as is seen in the clumps (c.f. Section \ref{subsec:enhancedB}).

Gas and magnetic pressures, and so total clump pressure, will increase with time. 
As a result, at certain point in time, the enhanced clump pressure stops further 
expansion of the I-front into the clump \citep[e.g.,][and references therein]{Ferland2008,Ferland2009}. 
For this to occur, a near pressure-equilibrium must be achieved between 
clumps and feedback, i.e., the total pressure 
within the clumps should be equal to or higher than the 
pressure imparted by the feedback from the H\,{\sc ii} region. 
Below we test this hypothesis. 

The pressure balance equation can be written as
\begin{equation}\label{eq:equipressure}
	P_{\mathrm{B}} + P_{\mathrm{turb}} + P_{\mathrm{Tg}}  =  P_{\mathrm{Te}} + P_{\mathrm{rad}},
\end{equation}
where the left-hand side (LHS) corresponds to the clump internal pressure,
$P_{\mathrm{clump}}$~$=$~$P_{\mathrm{B}}$ + $P_{\mathrm{turb}}$ + $P_{\mathrm{Tg}}$
(e.g., equation 14 of \citealt{Miaoetal2006}), which is the combination of magnetic ($P_{\mathrm{B}}$), turbulent (or non-thermal; $P_{\mathrm{turb}}$),
gas thermal (or kinetic; $P_{\mathrm{Tg}}$) pressures. The combination of the last two components can be treated as the
total molecular gas pressure ($P_{\mathrm{mol}}$) in the clumps, i.e., $P_{\mathrm{turb}}$ + $P_{\mathrm{Tg}}$~$=$~$P_{\mathrm{mol}}$.

Molecular gas pressure ($P_{\mathrm{mol}}$) is estimated using the following relations \citep[e.g.,][]{LiuTieetal2017}
\begin{equation}
P_{\mathrm{mol}} = n_{\mathrm{H_{2}}}\,k\,T_{\mathrm{eff}}
\end{equation}

and 

\begin{equation}
T_{\mathrm{eff}} = \frac{{C_{\mathrm{eff}}}^{2}\,m_{\mathrm{H}}\,\mu_{\mathrm{H}}}{k}.
\end{equation}

Using the $C_{\mathrm{eff}}$ values measured in Section \ref{subsubsec:stability_and_critical_mass}), 
the estimated $T_{\mathrm{eff}}$ values are 188$\pm$10 K and 148$\pm$11 K for clumps 1 and 2. 
Finally, $P_{\mathrm{mol}}$ is derived to be $(13\pm2)\times10^{-10}$ dyn cm$^{-2}$ for clump 1 and  $(2.7\pm0.5)\times10^{-10}$ dyn cm$^{-2}$ for clump 2.

The right-hand side (RHS) in Equation \ref{eq:equipressure} corresponds to 
feedback pressure, $P_{\mathrm{fb}}$~$=$~$P_{\mathrm{Te}}$ + $P_{\mathrm{rad}}$, due to the combination of 
the thermally ionized medium (from electron temperature; $P_{Te}$) and radiation ($P_{rad}$) components. 

$P_{\mathrm{clump1}}$ and $P_{\mathrm{fb1}}$ are estimated to be 41$\times$10$^{-10}$ and 8$\times$10$^{-10}$ dyn cm$^{-2}$  
for clump 1. Similarly, $P_{\mathrm{clump2}}$ and $P_{\mathrm{fb2}}$ are estimated to be 4.4$\times$10$^{-10}$ and 5.1$\times$10$^{-10}$ dyn cm$^{-2}$   
for clump 2. These parameters, holding to the relations $P_{\mathrm{clump1}}$~$>$~$P_{\mathrm{fb1}}$ and $P_{\mathrm{clump2}}$~$\simeq$~$P_{\mathrm{fb2}}$, suggest that 
clump 1 stops further expansion of the ionized region, whereas feedback pressure is nearly in equilibrium with the pressure in clump 2. 

Our analyses show that magnetic pressure dominates over thermal pressure (at least in clump 1; Section \ref{subsec:bpthermp}), 
and that B-fields within the clumps situated at the waist of the H\,{\sc ii} region can constrain the paths of I-fronts to blow away from the filament ridge. In RMHD simulations \citep{MackeyLim2011b} initially stronger B-fields are shown to 
confine the photoevaporation flow into a bar-shaped, dense, ionized ribbon which shields the I-front.
These features are observed in both the clumps of S201, as is clear from the PACS/70$\mu$m image
shown in Figure \ref{fig:polvecmaps}. Therefore, combined contributions from the anisotropic expansion of the I-front,
additional anisotropic pressure introduced by the B-fields in the primordial filament,
and the enhanced B-fields in the clumps would result in the formation of bipolar H\,{\sc{ii}} regions \citep[e.g.,][]{Deharvengetal2015,Samaletal2018}.

Furthermore, enhanced B-fields not only guide the ionized gas and aid the formation of 
bipolar H\,{\sc ii} regions but also shield the clumps from erosion. 
These signatures imply that the enhanced B-fields and underlying turbulence at the clump centers 
counteract gravitational collapse and hence delay the evolution of the clumps. 
Based on NIR polarimetry, \citet{ChenZhiweietal2017} have found that the B-field strength in 
a shell, N4, has been enhanced due to magnetically frozen-in gas being swept up into the expanding shell. As a result, the fragmented clumps 
in N4 remain in a magnetically subcritical state, indicating that the
B-field is the dominant force, stronger than gravity. Similarly, based on SCUBAPOL2 observations, 
\citet{Pattleetal2018} have probed B-fields in the denser parts of the `Pillars of creation' in M16 and found that 
initially B-fields are swept up, becoming aligned parallel to the pillars and later due to gas compression the B-field become stronger; and hence govern the evolution and longevity of the pillars.  

\subsection{Limitations of the current study}\label{subsec:limitations}

Due to limited sensitivity 
(mean rms noise in 850$\mu$m Stokes I is $\sim$5 mJy/beam with a bin size of 12$\arcsec$) 
achieved in our study and the relatively small field-of-view ($3\arcmin$ 
diameter with uniform sensitivity) of SCUBAPOL2 
observations, we could not probe B-fields in the clumps on the far side of the H\,{\sc ii} region. 
We also note that because of the small number of measurements made in clump 2, 
a reliable B-field strength has not been derived from ACF analysis. 

According to \citet{MackeyLim2011b}, initial weak (18 $\mu$G) and intermediate-strength (53 $\mu$G) 
B-fields that are initially oriented perpendicular to the I-front are swept into alignment with the 
pillar during its dynamical evolution, consistent with B-field observations in 
M16 \citep{Sugitanietal2007,Pattleetal2018}. Based on SCUBAPOL observations towards S106, 
\citet{Valleeetal2005} suggested that at large scales B-fields are roughly oriented along 
the north-south direction around the bipolar H\,{\sc ii} region, 
but close to central region near the IR star the B-fields are twisted into a 
toroidal morphology. Due to a lack of observations 
over an extended area, we could not examine these scenarios in S201. 

Based on MHD simulations of 
the evolution of sheet-like clouds due to mechanical stellar feedback 
from a single massive star (due both to stellar winds and to a supernova explosion),  
\citet{Wareingetal2017} showed that B-fields tend to follow a bipolar bubble-like structure 
similar to B-fields observed in RCW57A using NIR polarimetry \citep{Eswaraiahetal2017}. 
In order to examine whether B-fields follow and connect the structures of 
the photodissociation regions (PDRs) of the clumps (this work) 
as well as bipolar cavity walls \citep[][]{Eswaraiahetal2017}, further observations 
probing B-fields towards similar targets with SCUBAPOL2 are desirable.

MHD simulations focusing on the time evolution of B-fields, turbulence, gravity, and thermal energies, 
and their impact on the formation and evolution of clumps and bipolar H\,{\sc ii} regions would 
also be valuable.

\section{Summary and conclusions}\label{sec:summary_conclusions}

We have performed the dust continuum polarization observations at 850 $\mu$m and probed the B-fields in the deeply embedded massive clumps (clump 1 and clump 2) 
located at the waist of the bipolar H\,{\sc ii} region S201 using the JCMT SCUBA-2/POL-2. In addition, we have utilized 
JCMT/HARP molecular lines ($^{13}$CO (3--2) and C$^{18}$O (3--2)) and VLA 21~cm radio data, 
respectively, to quantify the turbulent and thermal energies in the region. In this work, we have derived 
various parameters such as B-fields, turbulence, gravity, and thermal pressures, and studied their interplay in the 
context of H\,{\sc ii} region-influenced star formation. The following are the main findings of our study: 

\begin{enumerate}

\item ~The morphological correlation between the orientations of B-fields and intensity gradients of 
	ionized gas (based on the VLA 21\,cm continuum) 
	suggests that B-fields are compressed and bent by the expanding ionization fronts from the H\,{\sc ii} region.

\item ~Compressed B-fields in clump 1 follow a bow-like morphology, while the degree of compression as well as of bending in B-fields is lower prominent in clump 2. The observed degree of compression, and of enhancement in the B-field strengths in clumps are 
in accordance with the amount of ionized medium interacting with the clump surfaces. 

\item ~We have employed structure function (SF) and auto-correlation function (ACF) analyses to derive the ratio of turbulent to ordered B-field in the clumps.
Using the velocity dispersion from C$^{18}$O data and column density derived from the POL2 Stokes I map, B-field strengths have been estimated by using 
the modified Davis-Chandrasekhar-Fermi relations. B-field strengths are found to be 266$\pm$32 $\mu$G (from ACF) for clump 1 and 65$\pm$6 $\mu$G (from SF) for clump 2. 

\item ~We find that magnetic pressure dominates in clump 1 and thermal pressure (due to ionized gas) dominates in clump 2. 
The amount of turbulent pressure lies between those of B-fields and those of ionized gas thermal energies in both clumps.

\item ~The comparison between clump internal pressure (magnetic, gas thermal, and non-thermal or turbulence) and 
feedback pressure (ionized gas thermal and radiation) imply that clump 1 has stopped further expansion of the H\,{\sc ii} region, while 
clump 2 maintains a near equilibrium with the feedback pressure. 

\item ~Virial analyses suggest that clump 1 is bound by its gravity and may collapse once it becomes unstable, whereas clump 2 is gravitationally unbound. We suggest that clump 2 may become bound in future if it progressively accumulates sufficient mass. 

\item ~Critical mass ratios reveal that the combined contribution from gas thermal energy, turbulence, and B-fields is not sufficient to counteract the gravity and hence is under collapse, whereas an opposite situation prevails in clump 2, in that its stability against collapse is maintained by these three factors. These results are consistent with the observed distribution of young stellar objects (YSOs), which suggests that star formation is ongoing in clump 1, while there is no star formation in clump 2. 

\item ~Feedback from H\,{\sc ii} region has the following consequences~--~(a)
causes the formation of clumps in the filament ridge, i.e., at the waist of the H\,{\sc ii} region, 
(b) enhances the B-field strength in the clumps and injects turbulence into the clumps, and 
(c) eventually the enhanced B-fields will be able to shield the clumps from erosion and so govern their stability, guiding the expanding I-fronts to be blown away from the filament ridge, and aiding in the formation of bipolar H\,{\sc ii} regions.

\end{enumerate}

\acknowledgments

We thank the anonymous referee for useful suggestions. 
This work is supported by the National Natural Science Foundation of China (NSFC) grant Nos. 11988101 and
11725313, and the International Partnership Program of Chinese Academy of Sciences grant No. 114A11KYSB20160008.
This work also supported by Special Funding for Advanced Users, budgeted and administrated by Center for 
Astronomical Mega-Science, Chinese Academy of Sciences (CAMS). 
C.E. and S.P.L. acknowledge the support from the Ministry of Science and Technology
(MOST) of Taiwan with grant MOST 106-2119-M-007-021-MY3. 
C.E. thanks David Berry for the help with JCMT SCUBA2-POL2 data reduction and analyses. 
A.Z. thanks the support of the Institut Universitaire de France. 
D.K.O. acknowledges the support of the Department of Atomic Energy,
Government of India, under project NO. 12-R$\&$D-TFR-5.02-0200. 
The James Clerk Maxwell Telescope is operated by the East Asian Observatory on behalf of
The National Astronomical Observatory of Japan; Academia Sinica Institute of Astronomy and Astrophysics;
the Korea Astronomy and Space Science Institute; Center for Astronomical Mega-Science.
Additional funding support is provided by the Science and Technology Facilities Council of the
United Kingdom and participating universities in the United Kingdom and Canada.

\facilities{JCMT (SCUBA-2/POL-2)}

\begin{table*}
	\centering
	\caption{Polarization measurements of S201 based on JCMT SCUBAPOL2 observations at 850 $\mu$m, along with the celestial coordinates of the pixels.}\label{tab:polmeasurements}
	\begin{tabular}{cccccccc}
\hline
RA (J2000) & Dec (J2000)   & $ P \pm \sigma$ &  $\theta \pm \sigma $ & $I \pm \sigma$  & $Q \pm \sigma$ & $U \pm \sigma$ & $PI \pm \sigma$ \\
  (degree) &  (degree)     & (\%)            &   (degree)            &  (mJy beam$^{-1}$) &  (mJy beam$^{-1}$) &  (mJy beam$^{-1}$) &  (mJy beam$^{-1}$) \\
 \hline
   45.771767 &    60.447989 &    13.4 $\pm$  5.1 &    164 $\pm$  10 &    38.6 $\pm$   3.0 &    -4.7 $\pm$   1.9 &     2.8 $\pm$   1.9 &     5.2 $\pm$   1.9 \\
   45.852871 &    60.457997 &    12.2 $\pm$  3.4 &    142 $\pm$  11 &    68.6 $\pm$   3.7 &    -2.1 $\pm$   3.3 &     8.4 $\pm$   2.2 &     8.4 $\pm$   2.3 \\
   45.839350 &    60.458000 &    17.0 $\pm$  1.2 &    153 $\pm$   2 &    86.1 $\pm$   2.8 &    -8.7 $\pm$   0.8 &    11.8 $\pm$   0.9 &    14.6 $\pm$   0.9 \\
   45.832592 &    60.458000 &    13.8 $\pm$  2.5 &    130 $\pm$   5 &   100.7 $\pm$   5.4 &     2.3 $\pm$   2.7 &    14.0 $\pm$   2.4 &    13.9 $\pm$   2.4 \\
   45.859637 &    60.461331 &     7.2 $\pm$  2.6 &     94 $\pm$  10 &   113.7 $\pm$   0.9 &     8.6 $\pm$   3.0 &     1.4 $\pm$   2.9 &     8.2 $\pm$   3.0 \\
   45.852875 &    60.461331 &     5.3 $\pm$  0.9 &    119 $\pm$   3 &   185.2 $\pm$   5.3 &     5.2 $\pm$   0.5 &     8.4 $\pm$   1.8 &     9.8 $\pm$   1.6 \\
   45.846112 &    60.461333 &     3.0 $\pm$  1.0 &    125 $\pm$   9 &   279.0 $\pm$   6.5 &     3.1 $\pm$   2.7 &     8.3 $\pm$   2.8 &     8.4 $\pm$   2.8 \\
   45.839354 &    60.461333 &     8.3 $\pm$  0.3 &    158 $\pm$   1 &   352.4 $\pm$   5.8 &   -21.4 $\pm$   0.9 &    20.1 $\pm$   0.8 &    29.4 $\pm$   0.8 \\
   45.832592 &    60.461333 &     6.0 $\pm$  0.3 &    144 $\pm$   4 &   302.8 $\pm$   4.6 &    -5.7 $\pm$   2.5 &    17.1 $\pm$   0.5 &    18.0 $\pm$   0.9 \\
   45.825829 &    60.461333 &     2.8 $\pm$  1.4 &    122 $\pm$  14 &   243.9 $\pm$   0.6 &     3.5 $\pm$   3.7 &     6.8 $\pm$   3.3 &     6.8 $\pm$   3.4 \\
\hline
\end{tabular}\\
A portion of the table is given here and in its entirety will be available online. 
\end{table*}

\begin{table*}[!ht]
\centering
	\caption{Various parameters for the two clumps of S201.}
	\label{tab:paramscl12}
	\begin{tabular}{clcc}\hline \hline
 	No &    Parameter &  clump 1 & clump 2      \\
         \hline
1 & clump dimensions$^{a}$ (fwhm$_{\mathrm {major}}$ pc, fwhm$_{\mathrm{minor}}$ pc, PA$\degr$) & 
0.334 $\pm$    0.004, 0.278  $\pm$  0.004, 83$\pm$6  &  0.401 $\pm$  0.005, 0.299$\pm$.005, 15$\pm$4 \\
  & clump dimensions (fwhm$_{\mathrm {major}}$ arcsec, fwhm$_{\mathrm{minor}}$ arcsec) & 34.43$\pm$0.92, 28.64$\pm$1.06 & 41.40$\pm$1.3,30.80$\pm$1.2 \\
2 & Effective radius ($R_{eff}$, pc)      &  0.129 $\pm$0.003 & 0.147$\pm$0.004 \\
		& Effective radius ($R_{eff}$, arcsec)    &  13.3$\pm$0.3 & 15.2$\pm$0.4 \\
3 & Gas number density (n(H$_{2}$; $\times$10$^{4}$) (cm$^{-3}$)  &  5.1$\pm$0.9  & 1.3$\pm$0.2 \\
4 & Mass (M$_{\sun}$)$^{a}$                  &  191$\pm$13    & 30$\pm$3 \\
5 & Dust temperature$^{d}$ ($T_{d}$; K)  & 27$\pm$2  & 29$\pm$1 \\
6 & Thermal velocity dispersion ($\sigma_{V_{T}}$; km s$^{-1}$) & 0.087$\pm$0.024  &  0.090$\pm$0.017 \\ 
7 & Non-thermal velocity dispersion ($\sigma_{V_{NT}}$; km s$^{-1}$) &  0.68$\pm$0.02 & 0.59$\pm$0.04 \\
8 & Turbulence pressure ($P_{turb}$; $\times$10$^{-10}$ dyn cm$^{-2}$    & 11$\pm$2     &  2.1$\pm$0.4 \\
9 & Thermal pressure ($P_{te}$; $\times$10$^{-10}$ dyn cm$^{-2}$)       & 8$\pm$2          &  4.7$\pm$0.3 \\
		10 & Radiation pressure  ($P_{rad}$; $\times$10$^{-10}$ dyn cm$^{-2}$) &  0.44          &  0.44 \\
11 & Sound speed (C$_{s}$; km s$^{-1}$) &  0.28$\pm$0.01  & 0.29$\pm$0.01 \\
12 & Effective sound speed (C$_{eff}$; km s$^{-1}$) &  0.74$\pm$0.02 &  0.66$\pm$0.04 \\  
13 & Effective temperature ($T_{eff}$; K) & 186$\pm$10 & 148$\pm$16\\
14 & Molecular gas pressure ($P_{mol}$; $\times$10$^{-10}$ dyn cm$^{-2}$)  & 13$\pm$2  & 2.7$\pm$0.5 \\
		\hline
       \multicolumn{3}{c}{From Structure function (SF) analyses}\\
       \hline
1 & $\left<\delta B^{2}\right>^{1/2}/\left<B_{o}\right>$     & 0.40$\pm$0.02   & 0.40$\pm$0.01 \\
2 & B-field strength (modified DCF) ($\mu$G)        & 147$\pm$15      & 65$\pm$6 \\
3 & B-field pressure ($P_{B}$; $\times$10$^{-10}$ dyn cm$^{-2}$)         & 9$\pm$2   & 1.7$\pm$0.3 \\
4 & $P_{B}/P_{turb}$                                 & 0.8$\pm$0.2     & 0.8$\pm$0.2  \\
5 & $P_{B}/P_{te}$                                & 1.0$\pm$0.3   & 0.4$\pm$0.1 \\
6 & $P_{turb}/P_{te}$                             & 1.3$\pm$0.4       & 0.4$\pm$0.1  \\
		\hline
        \multicolumn{3}{c}{From Auto-correlation function (ACF) analyses}\\
\hline
1 & $\left<\delta B^{2}\right>/\left<B_{o}^{2}\right>$     & 0.19$\pm$0.03   & 0.71$\pm$0.72 \\
2 & ($\left<\delta B^{2}\right>/\left<B_{o}^{2}\right>$)$^{1/2}$    & 0.43$\pm$0.04   & 0.84$\pm$0.43 \\
3 & Turbulent correlation length ($\delta$ in arcsec)     & 13$\pm$3  & 7$\pm$4    \\
4 & Coefficient ($a_{2}^{'}$; $\times$ $10^{-6}$)         & 41$\pm$13     &  6$\pm$14       \\
5 & B-field strength (modified DCF) ($\mu$G)                    & 266$\pm$32      & 61$\pm$31 \\
6 & B-field pressure ($P_{B}$; $\times$10$^{-10}$ dyn cm$^{-2}$)& 28$\pm$7    &  1.5$\pm$1.5 \\
7 & $P_{B}/P_{turb}$                                            & 2.6$\pm$0.8  & 0.7$\pm$0.7 \\
8 & $P_{B}/P_{te}$                                            &  3$\pm1$      & 0.3$\pm$0.3 \\
\hline
\end{tabular}\\
	\tablecomments{$^{a}$ Based on the CASA two-dimensional Gaussian fit assuming elliptical geometries for the clumps. 
	Obtained Gaussian sigma (radii) values are converted to FWHM using the relation FWHM~$=$~$\sqrt{8ln2}\sigma$.}
	\tablecomments{$^{b}$ By integrating the column densities, based on the {\it Herschel} dust temperature map \citep{Deharvengetal2012}, 
	within the 10$\sigma$ 850 $\mu$m Stokes I contour.}
	\tablecomments{$^{c}$ Using JCMT/HARP C$^{18}$O data from JCMT/HARP.}
	\tablecomments{$^{d}$ Based on dust temperature map from {\it Herschel} images \citep{Deharvengetal2012}}. \\
\end{table*}

\begin{table*}
        \centering
	\caption{Gaussian fit parameters ($T_{\mathrm{(b,p)}}$, $V_{\mathrm{LSR}}$, and $\sigma_{\mathrm{V_{LSR}}}$) for clumps 1 and 2
        based on $^{13}$CO(3--2) and C$^{18}$O(3--2) data from JCMT/HARP.}\label{tab:gaussfitresults}
        \begin{tabular}{ccccc}
\hline
		Clump & Spectral line & $T_{\mathrm{(b,p)}}$ (K) & $V_{\mathrm{LSR}}$ (km s$^{-1}$) & $\sigma_{\mathrm{V_{LSR}}}$ (km s$^{-1}$)\\
\hline
1 & $^{13}$CO(3--2) & 12.36 $\pm$ 0.07 & $-$40.70 $\pm$ 0.01 & 1.05 $\pm$ 0.01 \\
1 & C$^{18}$O(3--2) &  3.65 $\pm$ 0.09 & $-$40.69 $\pm$ 0.02 & 0.69 $\pm$ 0.02 \\
2 & $^{13}$CO(3--2) &  6.28 $\pm$ 0.05 & $-$40.22 $\pm$ 0.01 & 1.06 $\pm$ 0.01 \\
2 & C$^{18}$O(3--2) &  1.15 $\pm$ 0.06 & $-$40.24 $\pm$ 0.04 & 0.60 $\pm$ 0.04 \\
\hline
        \end{tabular}\\
	$T_{\mathrm{(b,p)}}$ = Peak brightness temperature \\
$V_{\mathrm{LSR}}$ = centroid velocity \\
$\sigma_{\mathrm{V_{LSR}}}$ = velocity dispersion \\
        \end{table*}

\bibliographystyle{aasjournal}
\bibliography{ref}

\appendix

\section{Intensity gradients}\label{sec:IngGrads_appendix}

In order to extract the directions of intensity gradients, we have used the VLA 21 cm/1.4 GHz continuum intensities of all the pixels in the map. 
Direction of gradients are computed for all the pixels except those at the edges, and when doing so we consider the intensities of adjacent pixels 
with respect to each central pixel. 
For the pixel ($\alpha_{i}$, $\delta_{j}$), where $i^{th}$ column and $j^{th}$ row 
correspond to RA ($\alpha$) and Dec ($\delta$), respectively, we estimate the intensity difference 
between the $(i+1)^{th}$ and $(i-1)^{th}$ pixels in RA, which is 
$\Delta I_{\alpha}$~$=$~$I_{\alpha_{i+1}} - I_{\alpha_{i-1}}$, and 
similarly the intensity difference between the $(j+1)^{th}$ and $(j-1)^{th}$ pixels in Dec, 
which is $\Delta I_{\delta}$~$=$~$I_{\delta_{j+1}} - I_{\delta_{j-1}}$. 
The position angle of the gradient $\theta^\prime_{IG}$ ($\degr$) for a pixel is then estimated using the relation
\begin{equation}
	\theta^\prime_{IG} = (180/\pi)\,\times\,\arctan\left[\frac{\Delta I_{\delta}}{\Delta I_{\alpha}}\right].
\end{equation}

In order to obtain the gradient directions ($\theta_{IG}$ values range between 0$\degr$ and 360$\degr$), 
the following corrections are employed on the above estimated $\theta^\prime_{IG}$ based on the values of $\Delta I_{\alpha}$ and 
$\Delta I_{\delta}$ 

\begin{equation}
\begin{aligned}
 \theta_{IG} & =
  \begin{cases}
	  \theta^\prime_{IG};~if~\Delta I_{\alpha}~>~0~\&~\Delta I_{\delta}~>~0, \\
          180 - \theta^\prime_{IG};~if~\Delta I_{\alpha}~<~0~\&~\Delta I_{\delta}~>~0, \\
          180 + \theta^\prime_{IG};~ if~\Delta I_{\alpha}~<~0~\&~\Delta I_{\delta}~<~0, \\
	  {\mathrm{and}}\\
          360 - \theta^\prime_{IG};~ if~\Delta I_{\alpha}~>~0~\&~\Delta I_{\delta}~<~0. \\
  \end{cases}\\
\end{aligned}
\end{equation}
Accordingly, the $\theta_{IG}$~$=$~0$\degr$ points towards celestial north and $\theta_{IG}$ 
increases towards the east. 
When we plot gradient orientations (c.f. Figure \ref{fig:diffpaBIG}) 
instead of directions, the $\theta_{IG}$ values that lie between $0\degr$ to 360$\degr$ are folded to obtain a range from 0$\degr$ to 180$\degr$. 
Similarly, while deriving the offset angle between $\theta_{B}$ and $\theta_{IG}$, an acute angle is computed that lies between 
0$\degr$ and 90$\degr$. 

\section{Optical depths and column densities traced by JCMT/HARP $^{13}$CO(3--2) and C$^{18}$O(3--2) data}\label{sec:tauco_appendix}

Since the excitation temperatures of both the lines $^{13}$CO and C$^{18}$O $J=3-2$ are unknown, we first assume that 
both lines are optically thin. Under this assumption, the peak brightness ratio of the $^{13}$CO to C$^{18}$O $J=3-2$ is expected to be similar 
to their abundance ratio. Assuming $\rm{[^{12}C/^{13}C] = 77}$, and $\rm{[^{16}O/^{18}O] = 560}$ \citep{Wilson1994}, 
the abundance ratio of [$\rm{^{13}CO /C^{18}O}$] is $\sim7.3$. As shown in Figure \ref{fig:13COandC18Ospectra}, 
however, the values of $T_{peak}^{13}/T_{peak}^{18}$ of clumps 1 and 2 are 3.3 and 4.3, respectively. This implies that 
our assumption of both lines being optically thin may be incorrect.

Following the assumption that the gas and dust are in local thermal equilibrium (LTE) conditions, 
the excitation temperatures of $^{13}$CO and C$^{18}$O emission from clumps 1 and 2 are approximated 
to be their dust temperatures as given in Table \ref{tab:paramscl12}, which are 27 and 29 K, respectively. 
According to equation 1 of \cite{Pineda2010}, and providing that the $^{13}$CO and C$^{18}$O emission fully fills the beam of 
the telescope, the optical depth can be expressed as
\begin{equation}
    \tau = -ln\{1 - \frac{T_{mb}}{T_0}[\frac{1}{(e^{T_0/T_{ex}}-1)^{-1} - (e^{T_0/T_{bg}} - 1)^{-1}}]\}
\end{equation}
where $T_{mb}$ is the main beam brightness temperature, $T_0 = h\nu_0/k$, $\nu_0$ is the rest frequency, and $T_{bg}$ is the 2.7 K 
cosmic microwave background radiation temperature. 
$T_{mb}$ and $\nu_0$ specifically refer to that of the $^{13}$CO or 
C$^{18}$O $J=3-2$ emission in this work. The upper-level column densities of the $^{13}$CO and C$^{18}$O 
molecules are related to the observed integrated intensities and optical depths of the 
two kinds of emission using equation 13 of 
\cite{Pineda2010}. Then the upper-level column density can be converted to the total column density using 
equation 17 with a 
partition function $Z = \Sigma(2J+1)exp[-hB_0(J+1)/(kT_{ex})]$, where J is the upper level and $B_0$ is the molecular rotational constant. 
The final column densities are related to the Plank constant $h$, Boltzman constant $k$, the Einstein A-coefficient $A_{UL}$, and the rest frequency $\nu_0$ of the $^{13}$CO and C$^{18}$O emission. 
The values of the constants used in this study are summarized in Table \ref{tab:cotaucold}, 
and are taken from the Splatalogue database (\url{https://www.cv.nrao.edu/php/splat/advanced.php}). 
The derived column densities of the $^{13}$CO and C$^{18}$O molecules are converted into column density of molecular hydrogen using the abundance ratios $\rm{[^{12}C/^{13}C] = 77}$, $\rm{[^{16}O/^{18}O]} = 560$, and $\rm{[H_2/^{12}CO]} = 1.1\times10^4$ \citep{Frerking1982}.

The optical depth and the column densities of the two clumps are derived from the averaged spectra within the extents of 
the ellipses shown in Figure \ref{fig:moment0maps}, and are tabulated in Table \ref{tab:cotaucold}. The optical depth $\tau^{13}$ of clump 1 is 1.01, 
which is moderately thick, while $\tau^{13}$ of clump 2 is well below 1. 
However, the C$^{18}$O emission of the two clumps is optically thin in both cases. 
Therefore, we choose the optically thin C$^{18}$O line emission to estimate the velocity dispersion of the two clumps.

\begin{table}
    \begin{center}
	    \caption{Constants and physical parameters of JCMT/HARP $^{13}$CO(3--2) and C$^{18}$O(3--2) data}
    \begin{tabular}{c|cccc}
        \hline
        \hline
        \multicolumn{5}{c}{ Related constants }\\
        \hline
		Emission line & $\nu_0$ & $\rm{B_0}$ & $\rm{A_{UL}}$ & Z\\
		&  (GHz) & (s$^{-1}$) & (s$^{-1}$) & \\
		\hline
		  $^{13}$CO $J=3-2$ & 330.59 GHz & $5.51\times10^{10}$ & $2.19\times10^{-6}$ &   10.54 (27K)/11.30 (29K) \\
		  C$^{18}$O $J=3-2$ & 329.33 GHz & $5.49\times10^{10}$ & $2.17\times10^{-6}$ &   10.59 (27K)/11.35 (29K)\\
		\noalign{}\hline
		\hline
		\multicolumn{5}{c}{ Physical parameters of the clumps }\\
		\hline
		Clump & $\tau^{13}$ & $\tau^{18}$ & $N_{H_2}^{13}$ & $N_{H_2}^{18}$\\
			&  & & (cm$^{-2}$) & (cm$^{-2}$) \\
		\hline
		  Clump 1  & 1.01 & 0.22 &  $2.1\times10^{23}$ & $2.0\times10^{23}$ \\
		  Clump 2  & 0.36 & 0.07 &  $7.9\times10^{22}$ & $5.0\times10^{22}$ \\
                 \hline

	    \noalign{}\hline
    \end{tabular}
    \label{tab:cotaucold}
    \end{center}
    \tablecomments{The upper part of the table gives the rest frequency $\nu_0$, Einstein A-coefficient $\rm{A_{UL}}$, rotational constant B$_0$, and the sum of the partition functions, Z, under different excitation temperatures of the $^{13}$CO and C$^{18}$O emission. The lower part shows the derived optical depth $\tau$ and H$_2$ column densities of clumps 1 and 2 determined using different emission lines. The superscripts 13 and 18 refer to the $^{13}$CO and C$^{18}$O emission, respectively.}
\end{table}

\clearpage

\end{document}